# On the Differential Topology of Expressivity of Parameterized Quantum Circuits

Johanna Barzen[0000-0001-8397-7973] and Frank Leymann[0000-0002-9123-259X]

University of Stuttgart, IAAS, Universitätsstr. 38, 70569 Stuttgart, Germany
{firstname.lastname}@iaas.uni-stuttgart.de

**Abstract.** Parameterized quantum circuits play a key role in quantum computing. Measuring the suitability of such a circuit for solving a class of problems is needed. One such promising measure is the expressivity of a circuit, which is defined in two main variants. The variant in focus of this contribution is the so-called dimensional expressivity which measures the dimension of the submanifold of states produced by the circuit.

Understanding this measure needs a lot of background from differential topology which makes it hard to comprehend. In this article we provide this background in a vivid as well as pedagogical manner. Especially it strives towards being self-contained for understanding expressivity, e.g. the required mathematical foundations are provided and examples are given.

Also, the literature makes several statements about expressivity the proofs of which are omitted or only indicated. In this article we give proofs for key statements from dimensional expressivity, sometimes revealing limits for generalizing them, and also sketching how to proceed in practice to determine this measure.

## 1. Introduction

Quantum algorithms are unitary operations transforming an initial state into another state that either represents directly the solution or that is a state that is measured and the measurement result lead to the solution of the problem [NC16]. Quantum algorithms are typically realized as quantum circuits: a composition of unitary operations at lower dimensions that represent the decomposition of the overall unitary transformation. These lower dimensional unitary operations are typically the operations supported by a concrete quantum computer. Sometimes, the unitary operations are parameterized: for example, these parameters are angles by which a state must be rotated. By varying the parameters between different executions of the circuit a solution is iteratively computed.

Such variational quantum algorithms (VQA) [C+21] and especially their encompassed parameterized quantum circuits (PQC) (Section 2.1) gain a lot of attention not only because their appropriateness for today's available computers (which are noisy [P18]), but also because parameterized quantum circuits represent applications in themselves (like quantum neural networks [H+21], [B+20]).

Whether or not a variational quantum algorithm can compute a solution of the problem addressed depends of the "suitability" of the parameterized quantum circuit it encompasses. Thus, having a measure of "suitability" in this context is of outmost importance. Such a measure is called *expressivity* of the parameterized quantum circuit and has been defined in two major variants:

- First, a parameterized quantum circuit associates with each parameter tuple a unitary matrix. The more of the unitary group is covered by the parameterized quantum circuit the better. This is because if circuits that solve the problem (i.e. unitary matrices) exist, then the unitary matrices generated by the VQA should include them. And the likelihood that this is achieved is higher the more of the unitary group is covered. The coverage of the unitary group is suggested to be assessed by means of how close the generated matrices are to be Haar random [SJA19]. This is what we call the "unitary approach"; it is described at a high level in section 2.3.1 and in more detail in section 4.8.
- The focus of this contribution is on the next variant of expressivity. While the first variant is about capturing a solving circuit, the second variant is about the ability to provide a solution itself. For this purpose the unitary matrices generated by the parameterized quantum circuit are immediately applied to a fixed initial state which results in another state, i.e. an element of the unit sphere in some Euclidian space $\mathbb{R}^n$. Any solution of the problem to be solved by the variational quantum algorithm is an element of this unit sphere. Thus, if the set of elements of the unit sphere covered by this process is a significant subset of the unit sphere the likelihood is high that a solution is reached. The amount of coverage of the unit sphere is assessed by means of the dimension of the submanifold generated by this process [F+21]. This is what we call the "state approach"; section 2.3.2 introduces this approach and provides corresponding definitions, while section 5 gives more details.

The issues with the "state approach" dealt with in this contribution is twofold: Readers of the corresponding existing quantum computing literature require a lot of background from differential topology. We compile and explain this background in detail and in a vivid as well as pedagogical manner; precise references to the matching mathematical literature are given. Furthermore, the corresponding quantum computing literature makes a bunch of statements and claims that are often not proven or proofs are only indicated. This makes reading and comprehending the literature difficult. We give detailed proofs for claims made in the context of the "state approach"; especially, we prove that the state approach produces in many situations locally immersed submanifolds and even embedded submanifolds (Section 5.2) such that speaking about "dimension" is meaningful at all. We also give a counter-example that these immersions and embeddings can in general not be extended to global immersions and embeddings (Section 5.3). Finally, we provide a "recipe" for determining how the local submanifolds can be enlarged (Section 5.4).

## 1.1. Structure of the Article

The article is structured as follows: Section 2 motivates the need for a measure of expressivity (Section 2.2) based on the structure and use of variational quantum algorithms (Section 2.1). The two major definitions of expressivity are given and discussed (Section 2.3). Section 3 proves that parameterized quantum circuits are differentiable maps, reminds the format of such circuits in many practical cases, and comments on the structure of the parameter space in practical situations.

Section 4 is the main pedagogical part: After proving a lemma on the existence of open chains in connected spaces (that is needed in Section 5.3), the concept of differentiable manifolds with boundaries is introduced, as well as some of their properties that are needed in the rest of this paper (Section 4.2). Section 4.3 discusses singularities which require special care in making claims or assumptions in our context. Next, differentiable maps and their differentials are discussed (Sections 4.4 and 4.5). Properties of maps with constant rank are reviewed and immersions as well as embeddings are discussed (Section 4.6). Submanifolds (both, immersed as well as embedded) are the subject of Section 4.7. In Section 4.8 the notion of the volume of manifolds is motivated by describing the concept of both, linear approximations of manifolds themselves (4.8.1) and "classical" volumes of parallelepipeds (4.8.2). Riemannian manifolds and their volume forms are introduced in Section 4.8.3. Curvature is revealed as the origin of non-uniform distribution of point sets on a manifold and the use of the Haar measure as solution allowing uniformly random distributions (Section 4.8.6). This allows us in Section 4.8.7 to introduce a precise definition of expressivity in the unitary approach.

Section 5 provides proofs supporting claims from the literature about the state approach. It is proven in Section 5.1 that parameterized quantum circuits induce local immersions. This implies that slices of the parameter space are locally mapped to immersed and even embedded submanifolds the dimensions of which is determined by the local rank of the map induced by the parameterized quantum circuit (Section 5.2). A usual technique to expand local properties to global ones is sketched in Section 5.3, and it is shown that this technique fails in our case, i.e. that local embeddings cannot be extended onto connected components. But a high-level proceeding is roughly sketched in Section 5.4 (together with an example) how to determine "large" areas of the parameter space that are mapped to an embedded submanifold by the parameterized quantum circuit. The contribution ends with a conclusion in Section 6.

## 2. The Notion of Expressivity

In the chapter we discuss the basic principles of variational quantum circuits (section 2.1), their origin (section 2.2.1), and the need for a metric to assess the "success probability" of such algorithms (section 2.2.2). Two approaches for such a metric - i.e. *expressivity* - are introduced in section 2.3.

## 2.1. Variational Quantum Algorithms

A *Variational Quantum Algorithm* (VQA) is a pair consisting of a parameterized quantum circuit (a.k.a. ansatz) and an optimizer ([C+21], [MR+16]). The ansatz prepares a state on a quantum computer that is measured [B+24]. The measurement result is passed to the optimizer that uses a problem-specific cost function to determine based on the measured values new parameters for the ansatz. The goal is to find parameters that produce a quantum state whose measurement result of which optimize the cost function.

Figure 1 depicts the structure and main ingredients of a variational quantum algorithm. The *parameterized quantum circuit* (PQC) $\mathscr{C}(p_1, \ldots, p_k)$ is given by a unitary operator $\mathscr{C}$ that depends on real parameters $p_1, \ldots, p_k \in \mathbb{R}$. The measurement of the quantum state produced by the quantum circuit is performed by means of problem-specific observables $\{O_m\}$. The cost function $C$ depends on the same parameters $p_1, \ldots, p_k$ typically by means of functions $f_i$ that have the measured expectation value $\langle O_i \rangle_{\mathscr{C}(p_1, \ldots, p_k)|s\rangle}$ as argument, where $\mathscr{C}(p_1, \ldots, p_k)|s\rangle$ is the state prepared by the ansatz $\mathscr{C}(p_1, \ldots, p_k)$ from the initial state $|s\rangle$:

$$C(p_1, \ldots, p_k) = \sum_i f_i\left(\langle O_i \rangle_{\mathscr{C}(p_1, \ldots, p_k)|s\rangle}\right) \quad (1)$$

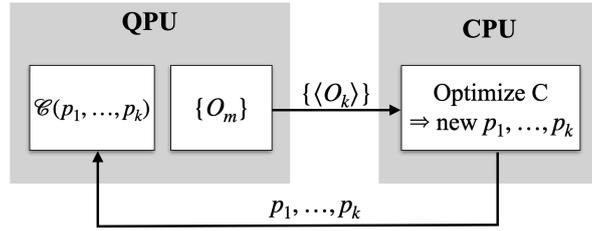

**Fig. 1**. Structure of a VQA

Next, an optimizer is used: it computes new parameters $p_1, \ldots, p_k$ that move the value of the cost function closer to an optimum. Any kind of optimizer can be used, e.g. gradient-based or derivative-free [NW06]. When the new parameters have been computed, they are feed into the parameterized quantum circuit $\mathscr{C}(p_1, \ldots, p_k)$, and the loop starts over again. The optimizer uses some termination condition like number of iterations performed or amount of improvements achieved by the new parameters. Once the termination criterion is satisfied the processing stops with the optimal parameters. The quantum circuit is run with the optimal parameters, and measurement of the prepared state in the computational basis represents the solution of the problem at hand.

A Quantum Neural Network (QNN) is also a parameterized quantum circuit [H+21]. Its cost function depends on training data in addition to other parameters. Once the optimal parameters have been determined (a.k.a. the QNN has been trained) the parameterized quantum circuit is run with the optimal parameters on new data input to perform the task of the QNN (e.g. classification).

Note, that the overall processing of determining the optimal parameters iterates between a quantum part performed on a QPU and a classical optimization part performed on a CPU. Such kind of processing is called *hybrid quantum-classical* processing, and the corresponding algorithm is called a hybrid quantum-classical algorithm (or just "hybrid" for short). In fact, most quantum algorithms and applications exploiting quantum resources are hybrid ([LB20], W+21]).

## 2.2. The Origin of Expressivity

The appropriateness of a variational quantum circuit should be able to be assessed. Given today noisy devices, the most fundamental criterion is that such an algorithm should be able to succeed at all (section 2.2.1). Next, varying the parameters of the circuit should be able to find an (approximately) correct solution of the problem: this means that either a unitary operation that solves the problem can be generated (section 2.3.1) or that the solution itself is determined by means of appropriate parameters (section 2.3.2).

*2.2.1. Suitability for NISQ*

Variational quantum algorithms have a lot of properties that are of principle interest, e.g. they are universal under certain conditions [MBZ20]. Also, some problems are inherently solved by variational quantum algorithms like quantum neural networks [B+20]. But the main operational aim behind variational quantum algorithms is to achieve an advantage or a utility of quantum computing in the NISQ era (Noisy Intermediate Scale Quantum). This era is roughly defined as the years during which quantum computers are still noisy and have only up to a few hundreds of qubits [P18]. "Noisy" means that the states of these qubits are stable only for a short period of time ("decoherence"), and the gates show small errors ("lack of fidelity"). In this era implementing quantum applications has a lot of issues to consider [LB20].

One key consequence of being noisy is that an algorithm must use a QPU only for a short period of time, i.e. less than the decoherence time $T_D$. A variational quantum algorithm, thus, strives towards an ansatz $\mathscr{C}(p_1, ..., p_k)$ and measurement $M$ via observables $\{O_m\}$ such that the time $T_\mathscr{C}$ that the ansatz is executed plus the time $T_M$ that the measurement requires is significantly smaller than the decoherence time:

$$T_\mathscr{C} + T_M \ll T_D \qquad (2)$$

*2.2.2. Finding Solutions*

The other fundamental aim is that the ansatz is capable in principle of finding a solution for the problem at hand. This capability can be assessed based on

    (i) the plethora of unitaries that an ansatz can produce, or
    (ii) the plethora of states that an ansatz can produce.

The first approach (that we call the "Unitary Approach" in what follows) considers the parameterized quantum circuit $\mathscr{C}(p_1, ..., p_k)$ as a map $\Omega$ that maps the parameter space $\mathbb{P}$ of all possible parameters $p_1, ..., p_k$ to the unitary group $U(n)$ based on the fact that each $p_1, ..., p_k \in \mathbb{P}$ results in a unitary map $\mathscr{C}(p_1, ..., p_k) \in U(n)$ (see [H+22]).

The second approach (called "State Approach" in the following) fixes an initial state $|s\rangle$ and uses for each $p_1, ..., p_k \in \mathbb{P}$ the unitary map $\mathscr{C}(p_1, ..., p_k) \in U(n)$ to result in an element $\mathscr{C}(p_1, ..., p_k)|s\rangle \in \mathbb{S}^{n-1}$, which is a map $\Lambda$ from the parameter space $\mathbb{P}$ to the unit sphere $\mathbb{S}^{n-1}$ (see [F+21]). Figure 2 depicts both approaches.

Then, a measure of the plethora of unitaries $\Omega(\mathbb{P}) \subseteq U(n)$ or a measure of the plethora of states $\Lambda(\mathbb{P}) \subseteq \mathbb{S}^{n-1}$, respectively, is referred to as the *expressivity* of the ansatz (or of the parameterized quantum circuit, respectively). Thus, in addition to finding an ansatz that satisfies equation (2), an ansatz should show a high expressivity. Finding such an ansatz is considered to be quite difficult (see [B+24], for example).

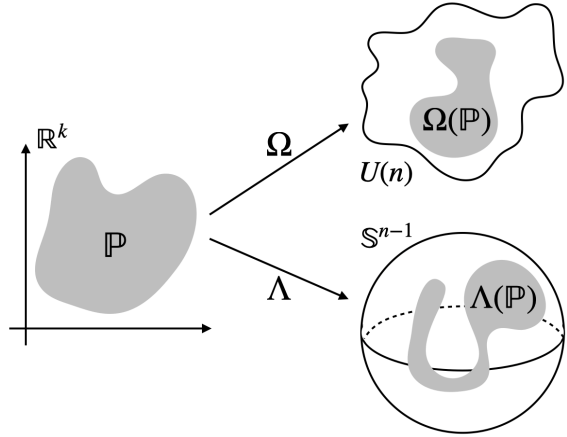

**Fig. 2**. Two Approaches to Assess Expressivity

### 2.3. Defining Expressivity

Expressivity (sometimes also called "expressibility") of an ansatz is measured, for example, by its ability to generate states that represent the Hilbert space well (see [SJA19]), i.e. to uniformly explore the entire Hilbert space (see [C+21]). In this section we define both approaches sketched in the section before more precisely. Note, that other definitions exist but they are often tailored toward specific questions (e.g. see [GED24]).

*2.3.1. The "Unitary Approach"*

Let $\mathbb{P} \subseteq \mathbb{R}^k$ be a parameter space and let $\mathscr{C}$ be the ansatz that depends on the parameters $p_1, ..., p_k \in \mathbb{P}$. The map $\Omega_\mathscr{C}$ is defined by $\Omega_\mathscr{C} : \mathbb{P} \to U(n)$ with $p_1, ..., p_k \mapsto \mathscr{C}(p_1, ..., p_k)$. Let $X$ be a problem to be solved by a quantum circuit, and let $\mathcal{S}(X) \subseteq U(n)$ be the set of unitary operators that solve the problem *X*, i.e. that produce a state that approximately optimizes the corresponding cost function.

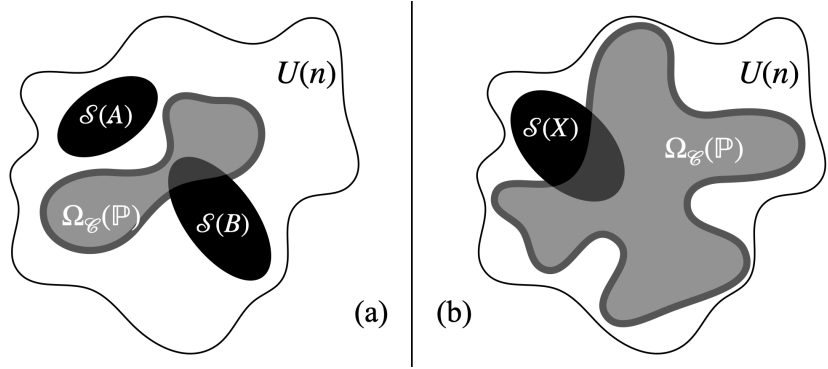

**Fig. 3**. (a) Completeness and (b) Expressiveness of an Ansatz: The Unitary Approach

If $\Omega_\mathcal{C}(\mathbb{P})$ contains an element of $\mathcal{S}(X)$ the ansatz $\mathcal{C}$ is called *complete* for the problem *X*. Part (a) of Figure 3 shows the solutions $\mathcal{S}(A)$ and $\mathcal{S}(B)$ of two problems *A* and *B*; the ansatz $\mathcal{C}$ is complete for problem *B* but not complete for problem *A*.

But the solutions $\mathcal{S}(X)$ of a problem *X* are not known in advance. Thus, the completeness of an ansatz cannot be assessed a priory. Consequently, the more of the unitary group $U(n)$ an ansatz covers (intuitively "the larger" $\Omega_\mathcal{C}(\mathbb{P}) \subseteq U(n)$ is) the higher is the likelihood that it will intersect the set of solutions $\mathcal{S}(X)$.

More precise: based of the fact that the unitary group $U(n)$ is a Riemannian manifold (see Note 10), a means to measure volumes is available on $U(n)$. Roughly, for an ansatz $\mathcal{C}$ the number $\text{vol}(\Omega_\mathcal{C}(\mathbb{P}))$ (i.e. the volume of $\Omega_\mathcal{C}(\mathbb{P})$) is called the *expressivity* of $\mathcal{C}$. If $\mathcal{C}$ and $\hat{\mathcal{C}}$ are two ansatzes, ansatz $\mathcal{C}$ is called more expressive than ansatz $\hat{\mathcal{C}}$ iff $\text{vol}(\Omega_\mathcal{C}(\mathbb{P})) \supset \text{vol}(\Omega_{\hat{\mathcal{C}}}(\mathbb{P}))$.

Note, that this definition before is not the definition used in the literature but it has the purpose of being descriptive: the exact definition of expressivity of an ansatz $\mathcal{C}$ based on the "unitary approach" compares its distribution of states from sampling with the collection of Haar-random states (see [SJA19], [H+21]). In section 4.8 we provide more background on this.

*2.3.2. The "State Approach"*

This approach is the one we focus on mostly, thus, we define it precise.

**Definition 1:** Let $\mathcal{C}$ be an ansatz depending on parameters $p_1, \ldots, p_k$. The set of all parameters $\mathbb{P} \subseteq \mathbb{R}^k$ allowed by $\mathcal{C}$ is called the *parameter space* of $\mathcal{C}$. Let $|\iota\rangle \in \mathbb{S}^{n-1}$ be a fixed chosen state, the so-called *initial state*. Then, the map $\Lambda_\mathcal{C}$ defined by $\Lambda_\mathcal{C} : \mathbb{P} \to \mathbb{S}^{n-1}$ with $p_1, \ldots, p_k \mapsto \mathcal{C}(p_1, \ldots, p_k)|\iota\rangle$ is called the *state map* of the parameterized circuit $\mathcal{C}$. ∎

$\Lambda_\mathcal{C}(\mathbb{P})$ is the set of all states reachable by means of the parameterized circuit $\mathcal{C}$. As proven in section 5.1, $\Lambda_\mathcal{C}(\mathbb{P})$ is under practical conditions locally an immersed submanifold (Lemma 17) of $\mathbb{S}^{n-1}$, and locally often even an embedded submanifold of $\mathbb{S}^{n-1}$ (Lemma 18). The reader is referred to sections 4.2 and 4.7 (or to textbooks like [G20] or [L13], respectively) for the definition of the terms "manifold" and "submanifold" (often, "manifolds" are assumed to be smooth, i.e. of class $C^\infty$, in what follows - which is without loss of generality according to Theorem 1).

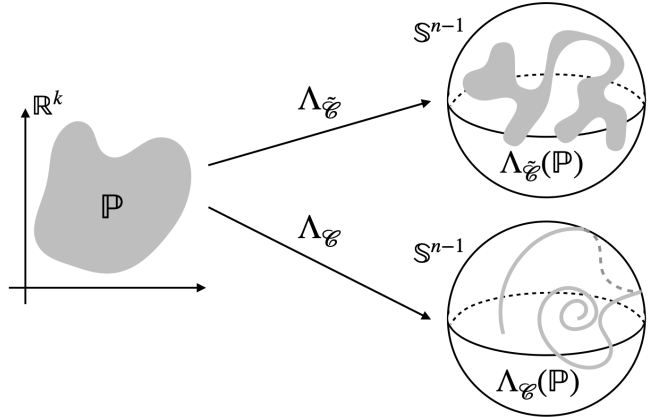

**Fig. 4**. Expressiveness of an Ansatz: The State Approach

Figure 4 depicts the images of the parameter space $\mathbb{P} \subseteq \mathbb{R}^k$ in the unit sphere $\mathbb{S}^{n-1}$ produced by two different ansatzes $\mathscr{C}$ and $\tilde{\mathscr{C}}$. In case a circuit produces a submanifold of $\mathbb{S}^{n-1}$ (immersed or embedded - see Section 5) this submanifold is called a *circuit manifold*, i.e. the circuit manifold is the set of states reachable by means of the parameterized circuit. In the figure, the circuit manifold $\Lambda_\mathscr{C}(\mathbb{P})$ has dimension 1, and the circuit manifold $\Lambda_{\tilde{\mathscr{C}}}(\mathbb{P})$ is of dimension 2. Intuitively, the dimension of a circuit manifold is a measure of the expressivity of a parameterized circuit [F+21]: the intuition is that a higher dimensional circuit manifold covers more of the inclosing manifold, thus has higher chance to meet the solution of a problem.

But this intuition is misleading: for example (see [L13], example 4.20), there is an immersed submanifold $L$ of the torus $\mathbb{T}^2$ which is dense in $\mathbb{T}^2$, but $\dim L = 1$. I.e. $L$ is arbitrary close to any point $p \notin L$. Thus, a circuit manifold of dimension 1 may approximate a solution with arbitrary precision (see Figure 5 (b) for a rough indication), while a "small" higher dimensional circuit manifold may stay "far away" from the solution (see Figure 5 (a)).

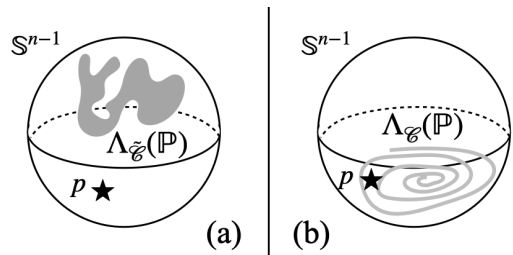

**Fig. 5**. Misleading Intuition of Dimensional

However, the dimensional expressivity allows to identify parameters of a parameterized circuit that are superfluous, i.e. that can be removed without affecting the capability of an ansatz to determine a solution (see [F+21] for the details).

## 3. Parameterized Quantum Circuits As Differentiable Maps

Expressivity depends on properties of parameterized circuits and their images in $U(n)$ or $\mathbb{S}^{n-1}$. The proofs of these properties are mostly sketched only in the literature. In what follows, these proofs are given in detail.

### 3.1. Proof of Differentiability

The most fundamental observation is that in many practical situations a parameterized quantum circuit is a differentiable map (even *smooth*, i.e. of class $C^\infty$). This can be seen by noting that each unitary can be represented by a set of 1-qubit operations and CNOTs, and by proving that these ingredient operations are smooth.

To begin with, we need the following fact from linear algebra:

**Note 1**: For each unitary matrix $A$ there exists a Hermitian matrix $H$ such that $A = e^{iH}$.

<u>Proof</u>: Each unitary is diagonalizable, i.e. there exists a $T \in U(n)$ such that $A = T \operatorname{diag}\left(e^{i\varphi_1}, \cdots, e^{i\varphi_n}\right) T^*$ with real numbers $\varphi_1, \ldots, \varphi_n \in \mathbb{R}$ (see [LM15] Theorem 18.13). Define the real diagonal matrix $\Phi := \operatorname{diag}(\varphi_1, \ldots, \varphi_n)$ and set $H := T\Phi T^*$. Then, $H$ is Hermitian because $H^* = (T\Phi T^*)^* = (T^*)^*\Phi^* T^* \stackrel{(a)}{=} T\Phi T^* = H$, where (a) is valid because $\Phi^* = \Phi$ for a real diagonal matrix $\Phi$. Then it is

$$\begin{aligned}
e^{iH} &= e^{iT\Phi T^*} \\
&\stackrel{(1)}{=} e^{iT\Phi T^{-1}} \\
&\stackrel{(2)}{=} T e^{i\Phi} T^{-1} \\
&= T e^{\operatorname{diag}(i\varphi_1, \ldots, i\varphi_n)} T^{-1} \\
&\stackrel{(3)}{=} T \operatorname{diag}(e^{i\varphi_1}, \ldots, e^{i\varphi_n}) T^{-1} \\
&= A
\end{aligned}$$

Hereby, (1) is valid because $T$ is unitary, i.e. $I = TT^*$, thus, $T^{-1} = T^*$. (2) is valid because $e^{SXS^{-1}} = Se^XS^{-1}$ for each $S \in \operatorname{GL}(n)$, which is seen as follows:

$$e^{SXS^{-1}} = \sum_{k=1}^{\infty} \frac{\left(SXS^{-1}\right)^k}{k!}$$

$$= \sum_{k=1}^{\infty} S \frac{X^k}{k!} S^{-1}$$

$$= S \left( \sum_{k=1}^{\infty} \frac{X^k}{k!} \right) S^{-1}$$

$$= S e^X S^{-1}$$

Finally, (3) is valid because

$$e^{\text{diag}(x_1,\ldots,x_n)} = \sum_{k=1}^{\infty} \frac{\text{diag}(x_1,\ldots,x_n)^k}{k!}$$

$$= \sum_{k=1}^{\infty} \frac{\text{diag}(x_1^k,\ldots,x_n^k)}{k!}$$

$$= \text{diag}(e^{x_1},\ldots,e^{x_n})$$

Thus, each unitary matrix is the matrix exponential of a Hermitian matrix. ∎

**Note 2**: Each parameterized 1-qubit operation $U : \mathbb{R} \supseteq I \to U(2)$ is smooth.

Proof: According to Note 1, each unitary matrix $U$ is a matrix exponential of a Hermitian matrix $H$: $U = e^{iH}$. Especially, for $I \subseteq \mathbb{R}$ and a parameterized 1-qubit operation $U : I \to U(2)$, $p \mapsto U(p)$, there exists a Hermitian matrix $H$ such that $U(p) = e^{ipH}$. Thus, $U(p)$ is a smooth operation. ∎

If the 1-qubit operation $U$ modifies the i-th qubit of an n qubit quantum register (i.e. all other qubits $j \neq i$ are left unchanged) it can be represented as the following tensor product:

$$T(p) = I \otimes \cdots \otimes I \otimes \overbrace{U(p)}^{i} \otimes I \cdots \otimes I \tag{3}$$

This is a smooth operation because the coefficients of the corresponding matrices are either constant (i.e. 0 or 1 in case the matrix $I$) or are the coefficients of the smooth 1-qubit operation $U(p)$. Finally, building the tensor product of matrices is a differentiable operation itself.

The same argument is used to prove:

**Note 3**: The controlled NOT operator with control qubit $r$ and target qubit $s$ $CX^{[r,s]} : \mathbb{H}^n \to \mathbb{H}^n$ on an n qubit quantum register is smooth.

Proof: The matrix $CX^{[r,s]}$ is a permutation matrix, i.e. its coefficients are constant $(CX^{[r,s]})_{ij} \in \{0,1\}$. Thus, the matrix is a smooth operator. ∎

A parameterized quantum circuit $\mathscr{C}(p_1, ..., p_k)$ has a parameter space $\mathbb{P} \subseteq \mathbb{R}^k$ (we will discuss properties of the topological structure of the parameter space in section 3.3), and for each parameter tuple $(p_1, ..., p_k) \in \mathbb{P}$ the resulting operator is a unitary: $\mathscr{C}(p_1, ..., p_k) \in U(n)$. Thus, a parameterized quantum circuit is a map

$$\mathscr{C} : \mathbb{P} \to U(n)$$
$$(p_1, ..., p_k) \mapsto \mathscr{C}(p_1, ..., p_k) \tag{4}$$

As a unitary, $\mathscr{C}(p_1, ..., p_k)$ is composed of (parameterized) 1-qubit operators (in the representation of equation (3)) and CNOT operators $CX^{[r,s]}$ (see [NC16], section 4.5.2). This composition is performed by multiplying the corresponding matrices. But matrix multiplication is smooth: if $(a_{ij}), (b_{st})$ are two matrices then their product is a matrix the coefficients of which are a sum of the products of coefficients of the factor matrices, i.e. $(a_{ij}) \cdot (b_{st}) = (\sum_j a_{ij} b_{jt})_{i,t}$, thus multiplying matrices is smooth. If the factor matrices of the matrix product are smooth (i.e. their coefficients $(a_{ij}), (b_{st})$ are smooth functions), the result is smooth again. Because the factors of the matrix product resulting in $\mathscr{C}(p_1, ..., p_k)$ are smooth parameterized 1-qubit operators and smooth CNOTs (see Note 2 and Note 3), $\mathscr{C}(p_1, ..., p_k)$ is smooth.

Let $(\mathscr{U}_{ij}(p_1, ..., p_k))_{i,j}$ be the matrix representation of $\mathscr{C}(p_1, ..., p_k)$. We choose an arbitrary but fixed initial state $|\iota\rangle = |\iota_1, ..., \iota_n\rangle \in \mathbb{H}^n$. Then the product $\mathscr{U}(p_1, ..., p_k)|\iota\rangle = (\sum_j \mathscr{U}_{ij}(p_1, ..., p_k) \cdot \iota_j)_i$ is smooth, i.e. the components of the resulting vector are smooth. This finally proves:

**Lemma 1**: Let $\mathscr{C} : \mathbb{P} \to U(n)$ be a parameterized quantum circuit with matrix representation $(\mathscr{U}_{ij}(p_1, ..., p_k))$ and let $|\iota\rangle \in \mathbb{S}^{n-1}$ be a fixed initial state. Then, both maps $\Omega_{\mathscr{C}} : \mathbb{P} \to U(n)$ with $(p_1, ..., p_k) \mapsto \mathscr{C}(p_1, ..., p_k)$ as well as $\Lambda : \mathbb{P} \to \mathbb{S}^{n-1}$ with $(p_1, ..., p_k) \mapsto \mathscr{U}(p_1, ..., p_k)|\iota\rangle$ are smooth. ∎

Figure 6 summarizes the ingredients of a parameterized quantum circuit: The circuit has an associated parameter space $\mathbb{P}$. The parameterized quantum circuit $\mathscr{C}$ itself assigns to each parameter tuple $(p_1, ..., p_k) \in \mathbb{P}$ a unitary operator $\mathscr{C}(p_1, ..., p_k) \in U(n)$ (we do not distinguish between the circuit $\mathscr{C}$ and its matrix representation $\mathscr{U}$). This unitary operator transforms an initial state $|\iota\rangle \in \mathbb{S}^{n-1}$ into another state $\mathscr{C}(p_1, ..., p_k)|\iota\rangle \in \mathbb{S}^{n-1}$. Thus, a parameterized quantum circuit $\mathscr{C}$ induces a smooth map $\Lambda : \mathbb{P} \to \mathbb{S}^{n-1}$ that maps its parameter space $\mathbb{P}$ into the unit sphere $\mathbb{S}^{n-1}$.

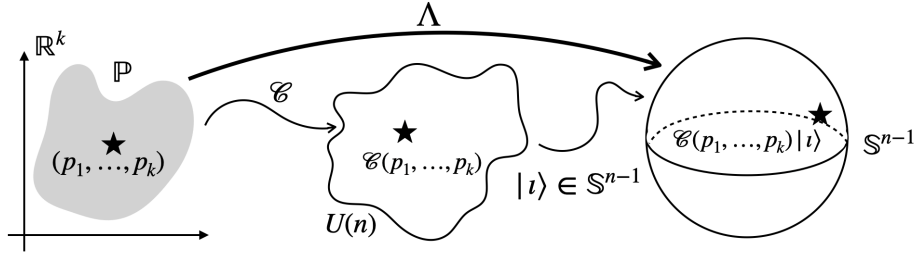

**Fig. 6**. Parameterized Quantum Circuit as a Smooth Map

### 3.2. Differentiability in Practice

In addition, an argument from practice can be used to see that a typical parameterized quantum circuit of a variational quantum algorithm is smooth. This is because often (see [C+21], [H+22]), a typical circuit has the format

$$\mathscr{C} : \mathbb{P} \to U(n)$$
$$(p_1, \ldots, p_k) \mapsto \prod_{j=1}^{k} e^{-ip_j H_j} W_j \quad (5)$$

with a set of fixed unitary operators $\{W_j\}$ and Hermitian operators $\{H_j\}$, i.e. $e^{-ip_j H_j}$ is a rotation of angle $p_j$ generated by $H_j$. The latter rotations a smooth, their product with a constant matrix are smooth, and, thus, the overall product is smooth. The parameter space $\mathbb{P}$ in this case is a Cartesian product of connected intervals in $\mathbb{R}$ which are the domains of the corresponding angles $p_j$.

### 3.3. Differentiability on Parameter Spaces

By now we did not comment on the structure of a parameter space $\mathbb{P}$ (except for the special case of section 3.2). Typically, a parameter space is not an arbitrary set in $\mathbb{R}^k$ but it is assumed to have suitable properties. Especially, $\mathbb{P}$ must support the notion of maps that are differentiable on $\mathbb{P}$ as their domain (see section 4.4). For example, for arbitrary sets $\mathbb{P} \subseteq \mathbb{R}^k$ a map $f : \mathbb{P} \to \mathbb{R}^n$ is differentiable, iff a function $F : U \to \mathbb{R}^n$ with $\mathbb{P} \subseteq U \subseteq_{\text{open}} \mathbb{R}^k$ exists that is differentiable in the ordinary sense and that fulfills $F|_\mathbb{P} = f$, i.e. $F$ restricted to $\mathbb{P}$ is identical to $f$ (in this case, $F$ is called a *differentiable extension* of $f$) - see Definition 5.

As another example, if $\mathbb{P}$ is a differentiable manifold (with or without boundary), differentiability is defined in every point $x \in \mathbb{P}$ by means of an appropriate chart around $x$ (see section 4.4 and [L13]).

In circuits like the ones in equation (5), $\mathbb{P}$ is a cartesian product of open intervals. This can slightly be generalized as follows (see section 4.2): $\mathbb{P} = \mathbb{P}_1 \times \ldots \times \mathbb{P}_k \subseteq \mathbb{R}^k$ with $\mathbb{P}_i \subseteq_{\text{Mf}} \mathbb{R}$ being a connected manifold (with or without boundary) of dimension one, $\dim \mathbb{P}_i = 1$. Such a manifold $\mathbb{P}_i$ is an open, closed or semi-closed interval. Thus, $\mathbb{P}$ is a hypercube where some of its faces may belong to the hypercube; note that not all of the intervals $\mathbb{P}_i$ may include boundary points in order to avoid singularities - see section 4.3. Thus, in general $\mathbb{P}$ is a manifold with boundary (section 4.2).

## 4. Facts from (Differential) Topology

In this section we prove that two points in a connected topological space are connected by an open chain (Lemma 2). The definition of manifolds with boundaries is given, as well as differentiability of maps. The concept of singularities is presented (section 3.3). Differentiability of maps, their differentials, and properties of maps of constant rank are summarized. Also, we remind the rank theorem which implies that immersions are locally injective, and we remind a condition under which injective immersions are embeddings. Submanifolds introduced and their properties are discussed. Section 4.8 introduces the volume form on manifolds as well as the Haar measure; this allows to present in section 4.8.7 more details about "unitary approach" (as sketched informally in section 2.3.1).

### 4.1. A Property of Connected Spaces

In this section we remind the notion of connectedness of a topological space (refer to the textbook [P22] for details). We prove a property of connected spaces (Lemma 2) that we need later on and that we did not find in the English literature (but in the German textbook [vQ01]).

**Definition 2:** A topological space $X$ is called *connected* if for two non-empty open sets $O_1, O_2 \subseteq_{\text{open}} X$, $O_1 \neq \emptyset$, $O_2 \neq \emptyset$ with $X = O_1 \cup O_2$ it follows that $O_1 \cap O_2 \neq \emptyset$. ∎

In other words, a topological space is connected if it cannot be split into two disjoint, non-empty open sets. The following is well-known (e.g. [P22], Proposition 9.8):

**Note 4**: Let $X$ be a connected topological space and let $Z \subseteq X$ be both, open and closed. Then $Z = \emptyset$ or $Z = X$. ∎

**Definition 3:** A family of open sets $\mathcal{U} = \{U_i\}_{i \in I}$, $U_i \subseteq_{\text{open}} X$ with $\bigcup_{i \in I} U_i = X$ is called an *open cover* of $X$. ∎

The proof of the following lemma is from [vQ01].

**Lemma 2**: Let $X$ be a topological space. $X$ is connected ⇔ For each open cover $\mathcal{U}$ of $X$ and two points $a, b \in X$ there are $\{U_1, \ldots, U_n\} \subseteq \mathcal{U}$ with:

(i)   $a \in U_1, a \notin U_i$ for $i \neq 1$
(ii)  $b \in U_n, b \notin U_i$ for $i \neq n$
(iii) $U_i \cap U_j \neq \emptyset \Leftrightarrow |i - j| \leq 1$

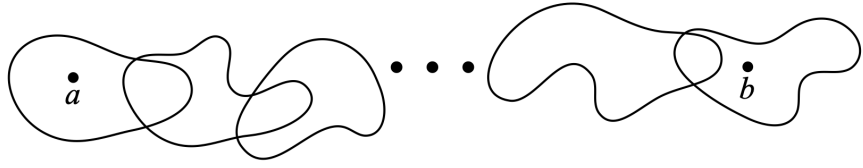

**Fig. 7**. Chain of Open Sets Connecting Two Points

An open cover $\mathcal{U}$ with the properties (i), (ii), (iii) for each pair of points $a, b \in X$ is called an *open chain* connecting the points $a, b \in X$ (see Figure 7).

Proof: "$\Leftarrow$" Let $X$ be disconnected. We have to show that there is an open cover of $X$ that does not satisfy conditions (i), (ii) and (iii). Because $X$ is not connected, we find open sets $O_1, O_2 \subseteq_{\text{open}} X$ with $O_1 \neq \emptyset$, $O_2 \neq \emptyset$, $O_1 \cap O_2 = \emptyset$ and $O_1 \cup O_2 = X$. Choose $a \in O_1$ and $b \in O_2$. The open cover $\mathcal{U} = \{O_1, O_2\}$ satisfies (i) and (ii) but not (iii).

"$\Rightarrow$" Let $\mathcal{U}$ be an open cover of $X$. Two points $a, b \in X$ are called $\mathcal{U}$-connected (in symbols $a \sim b$) if there exists $\{U_1, ..., U_n\} \subseteq \mathcal{U}$ with the properties (i), (ii), and (iii).

Then, "$\sim$" is an equivalence relation: reflexivity and symmetry are obvious. Transitivity is seen as follows: for $a, b, c \in X$ with $a \sim b$ and $b \sim c$ choose a chain of open sets $\{U_1, ..., U_n\} \subseteq \mathcal{U}$ connecting $a, b$ and choose a chain of open sets $\{V_1, ..., V_m\} \subseteq \mathcal{U}$ connecting $b, c$.

Set $r := \min\{i \in \{1, ..., n\} \mid \exists 1 \leq j \leq m : U_i \cap V_j \neq \emptyset\}$ and $s := \max\{j \in \{1, ..., m\} \mid U_r \cap V_j \neq \emptyset\}$. Then, $\{U_1, ..., U_r, V_s, ..., V_m\} \subseteq \mathcal{U}$ is a chain of open sets with properties (i), (ii), (iii) connecting $a, c$, i.e. $a \sim c$.

Let $[x] \subseteq X$ be a $\sim$-equivalence class. If we show that $[x] = X$ then we are done. For $y \in [x]$ and a chain of open sets $\{U_1, ..., U_n\} \subseteq \mathcal{U}$ connecting $x, y$ it is $U_n \subseteq_{\text{open}} X$ by definition, and it is $U_n \subseteq [x]$ because each $z \in U_n$ is $\mathcal{U}$-connected with $x$. Thus, $[x] \subseteq_{\text{open}} X$. This implies that $X \setminus [x] = \bigcup_{y \notin [x]} [y] \subseteq_{\text{open}} X$, thus $[x] \subseteq_{\text{closed}} X$.

Thus, $[x] \subseteq X$ is open and closed in $X$, but $[x] \neq \emptyset$. Note 4 before implies $[x] = X$. ∎

### 4.2. Differentiable Manifolds

The definition of a manifold with boundaries is given (the seminal textbook [H76] provides many details and proofs, and [L13] is a more modern treatment of the subject). For this purpose we need the definition of the n-dimensional real half-space $\mathbb{R}^n_+ := \{x \in \mathbb{R}^n \mid x_n \geq 0\}$; it is $\partial \mathbb{R}^n_+ = \{x \in \mathbb{R}^n \mid x_n = 0\} \cong \mathbb{R}^{n-1}$ the *boundary* of $\mathbb{R}^n_+$.

Note, that in the context of differentiable manifolds, topological spaces are considered to be Hausdorff spaces and second countable. A *Hausdorff space* requires that any two different points of the space have disjoint neighborhoods; the set of all neighborhoods of a point $p$ is denoted by $\mathfrak{U}_p$. A space is *second countable* iff it has a countable basis, i.e. every open set of the space is the union of a subset of the basis. See [L24] for the detailed definitions of these terms.

**Definition 4:** Let $M$ be a Hausdorff and second countable topological space. If for each point $x \in M$ there exists an open neighborhood $U \subseteq M$ and a homeomorphism $\varphi : U \to \varphi(U) \subseteq_{\text{open}} \mathbb{R}^n_+$, then $M$ is called a *topological manifold with boundary* of dimension $n$ (in symbol: $\dim M = n$). The pair $(U, \varphi)$ is called a *chart of $M$ around $x$*. A set of charts $\mathfrak{A} = \{(U_i, \varphi_i) \mid i \in I\}$ with $\bigcup_{i \in I} U_i = M$ is called an *atlas of $M$*.

A point $x \in M$ with $\varphi(U) \subseteq_{\text{open}} \mathbb{R}^n$ is called *interior point*, and a point $x \in M$ with $\varphi(x) \in \partial \mathbb{R}^n_+$ is called *boundary point*. The set of boundary points of $M$ is called the *boundary of $M$*, denoted by $\partial M$.

For two intersecting charts $(U_i, \varphi_i), (U_j, \varphi_j) \in \mathfrak{A}$, i.e. charts with with $U_i \cap U_j \neq \emptyset$ the map $\varphi_i \circ \varphi_j^{-1} : \varphi_j(U_i \cap U_j) \to \varphi_i(U_i \cap U_j)$ is called the *transition function* between the charts.

$M$ is called a *differentiable manifold of class $C^r$* ($C^r$-manifold for short) if all transition functions are differentiable of class $C^r$; the corresponding atlas $\mathfrak{A}$ is called $C^r$-atlas. For $r = \infty$, the manifold (and the atlas) is called *smooth*.  ∎

In Figure 8 $p$ and $q$ are interior points while $r$ is a boundary point.

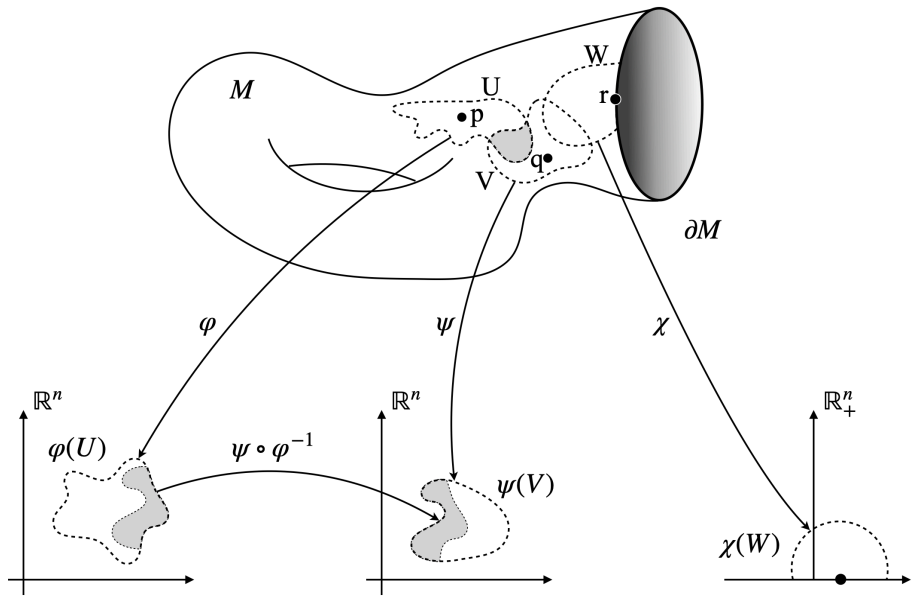

**Fig. 8**. Charts and Transition Functions of a Differentiable Manifold With Boundary

A $C^r$-manifold may have several $C^r$-atlases. For example, both, $\mathfrak{A}_1 = \{(\mathbb{R}^k, \text{id})\}$ as well as $\mathfrak{A}_2 = \{(\mathbb{R}^k_{x_k > -1}, \text{id}), (\mathbb{R}^k_{x_k < 1}, \text{id})\}$ are $C^r$-atlases (for any $r$) of $\mathbb{R}^k$, and so is $\mathfrak{A}_1 \cup \mathfrak{A}_2$. In general, when adding a chart $(U, \varphi)$ to a given $C^r$-atlas $\mathfrak{A}$ and the resulting atlas $\mathfrak{A} \cup \{(U, \varphi)\}$ is again a $C^r$-atlas, $(U, \varphi)$ is said to be *compatible* with $\mathfrak{A}$. Adding all compatible charts to $\mathfrak{A}$ results in the (unique) maximal atlas $\overline{\mathfrak{A}}$ (that contains $\mathfrak{A}$). A maximal $C^r$-atlas is called a $C^r$-*differentiable structure* of the corresponding manifold. Any $C^r$-atlas $\mathfrak{A}$ determines a unique $C^r$-differentiable structure (see [L13], Proposition 1.17). Thus, we can assume that the atlases of a manifold are differentiable structures.

The following is often used, and its proof can be found in [H76].

**Lemma 3**: Let $M$ be a $C^r$-manifold with boundary, $\dim M = n$. Then, $\partial M \subseteq_{\text{closed}} M$ and $\partial M$ is a $C^r$-manifold without boundary, $\dim \partial M = n - 1$. ∎

Also, any open subset of a manifold is again a manifold:

**Note 5**: (a) $\mathbb{R}^k$ is a smooth manifold without boundary.
(b) For a $C^r$-manifold $M$ any open subset $S \subseteq_{\text{open}} M$ is a $C^r$-manifold.
(c) The interior $M \setminus \partial M$ of a manifold with boundary is a manifold without boundary.

Proof: (a) $\mathfrak{A} = \{(\mathbb{R}^k, \text{id})\}$ is a smooth atlas that shows that $\mathbb{R}^k$ is a manifold without boundary.

(b) If $(U, \varphi)$ is a chart of $M$ around $x \in S$, then $x \in U \cap S$ and $\varphi|_{U \cap S}$ is a homeomorphism. Thus, $(U \cap S, \varphi|_{U \cap S})$ is a chart of S around $x$. Also, restrictions of transition functions are of the same differentiability class than the original functions.

(c) It is $M \setminus \partial M \subseteq_{\text{open}} M$, thus, the claim follows from (b). ∎

There are several ways how manifolds can be constructed from existing manifolds, e.g. by building their sums, quotients, and products (see [L13]). In our context, the product of manifolds is of interest: if $\{M_1, \ldots, M_k\}$ are $C^r$-manifolds (without boundary) and $\dim M_i = n_i$, then their cartesian product $M_1 \times \cdots \times M_k$ is another $C^r$-manifold (without boundary). For example:

- The unit sphere $\mathbb{S}^1 \subseteq \mathbb{R}^2$ is a smooth manifold. Thus, $\mathbb{T}^n := \mathbb{S}^1 \times \cdots \times \mathbb{S}^1$ (the product of n copies of $\mathbb{S}^1$) is a new smooth manifold called n-dimensional torus.

- An open interval $]a_i, b_i[ \subseteq_{\text{open}} \mathbb{R}$ (for $1 \leq i \leq n$) is a smooth manifold. Thus, the open n-dimensional cuboid $Q^n := ]a_1, b_1[ \times \cdots \times ]a_n, b_n[ \subseteq \mathbb{R}^n$ is a smooth manifold.

For manifolds with boundaries building their products is a bit less straightforward. In general, there is an extensive theory behind this (see [LM24]). In our context, the following facts suffice:

1. Let $\{M_1, \ldots, M_k\}$ be $C^r$-manifolds (without boundary), $\dim M_i = n_i$, and let $N$ be a $C^r$-manifold with boundary, $\dim N = n$. Then, $M_1 \times \cdots \times M_k \times N$ is a manifold with boundary, $\dim(M_1 \times \cdots \times M_k \times N) = n_1 + \cdots + n_k + n$, and $\partial(M_1 \times \cdots \times M_k \times N) = M_1 \times \cdots \times M_k \times \partial N$ (see [L13], Proposition 1.45).

2. Thus, with $]a_i, b_i[ \subseteq_{\text{open}} \mathbb{R}$ for $1 \leq i \leq n$ and $[a, b] \subseteq_{\text{closed}} \mathbb{R}$, the product $Q := ]a_1, b_1[ \times \cdots \times ]a_n, b_n[ \times [a, b] \subseteq \mathbb{R}^{n+1}$ is a manifold with boundary

$$\partial Q = ]a_1, b_1[ \times \cdots \times ]a_n, b_n[ \times \{a\} \cup ]a_1, b_1[ \times \cdots \times ]a_n, b_n[ \times \{b\}$$ (for an example see Figure 9).

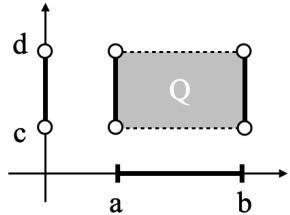

**Fig. 9**. A Product Manifold Q With Boundary

### 4.3. Singularities

In Figure 9, the boundary of the manifold Q consists of the left and right edges without their endpoints. This is an implication of the fact that a boundary of a manifold with boundary is a manifold without boundary (see Lemma 3). If the endpoints of the left or right edges would be included, they would become the boundary of the manifolds consisting of the left or right edges: contradiction.

This is a general phenomenon: manifolds exclude *singularities*, i.e. non-differentiable structures. A vertex of a rectangle is an example of such a singularity: Part (a) in Figure 10 depicts the rectangle $Q$ and one of its vertices $x$. Part (b) focusses on a neighborhood of $x$ in $Q$. Finally, part (c) extracts the boundary of this neighborhood, moves $x$ to the origin, and rotates the boundary by 45°; movements and rotations are smooth maps, i.e. the resulting graph in (c) is diffeomorphic to the part of the boundary in (b). Obviously, the graph in (c) is the graph of the absolute function. Assume that there is a chart $(U, \varphi)$ of the boundary around $x$ with $\varphi(x) = 0$. Then $\varphi$ is a diffeomorphism (see Note 6 below), thus, $\varphi^{-1}$ is differentiable, but $\varphi^{-1}$ is the absolute function which is known to be not differentiable at 0. This contradiction shows that $x$ is a singularity, i.e. vertices must not be part of a rectangular "shape" in order to be a manifold.

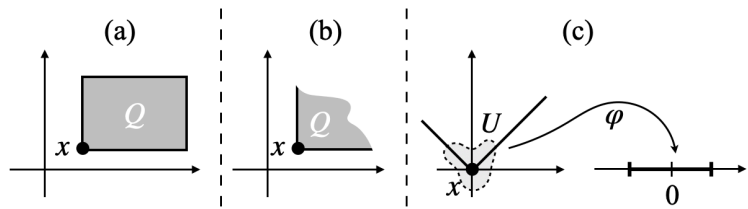

**Fig. 10**. Vertices in Rectangles are Singularities

Many other kinds of singularities exist. For example (see Figure 11): Part (a) of the figure shows a "cusp" which is a curve that is not differentiable at the point $x$. An intersection at point $y$ in part (b) is not differentiable. Also, an edge of a cuboid (see part (c) of the figure) consists of nodes each of which is a singularity.

The intersection point in part (b) is even already a topological singularity, i.e. no reference to differentiability is needed to recognize this: Any point different from $y$ is contained in a neighborhood that is homeomorphic to an open interval in $\mathbb{R}$. Thus, if a chart $(U, \varphi)$ around $y$ would exist, U could be chosen to be connected. Then, $\varphi(U) \subseteq \mathbb{R}$ is connected. $U \setminus \{y\}$ consists of four connected components. Deleting a single point from a connected subset of the real line results in two connected components, i.e. $\varphi(U) \setminus \{\varphi(y)\} = \varphi(U \setminus \{y\})$ consists of two connected components - contradiction because $\varphi(U \setminus \{y\})$ consists of four connected components.

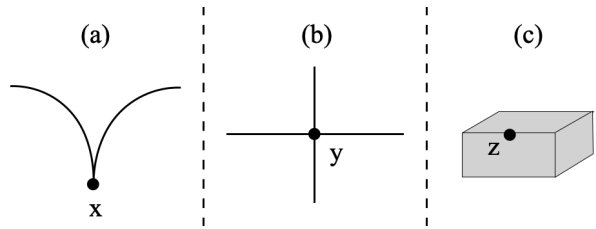

**Fig. 11**. More Singularities

Many of the results achieved in differential topology of manifolds are not valid in the presence of singularities. This is why special care must be taken when claiming that geometric objects are manifolds (with or without boundary): it must be proven that they are manifolds to avoid singularities. Singularities are extensively studied (see [A84], [BL75], or [J16] for example).

### 4.4. Differentiable Maps

In Figure 8, $\chi(W)$ is not an open set in $\mathbb{R}^n$, thus, the ordinary definition of differentiability (that is typically defined for open sets) does not apply. Consequently, the notion of differentiability is extended to functions with a domain that is an arbitrary subset of $\mathbb{R}^n$.

**Definition 5:** Let $A \subseteq \mathbb{R}^n$ be an arbitrary set and let $f : A \to \mathbb{R}^k$ be a map. $f$ is called *differentiable of class $C^r$*, iff a map $F : U \to \mathbb{R}^k$ with $A \subseteq U \subseteq_{\text{open}} \mathbb{R}^n$ exists that is differentiable of class $C^r$ in the ordinary sense and that fulfills $F|_A = f$, i.e. $F$ restricted to $A$ is identical to $f$. $F$ is called a *differentiable $C^r$-extension* of $f$. ∎

Based on this we can define differentiable maps between manifolds (see also Figure 12):

**Definition 6:** Let $M$ and $N$ be two $C^r$-manifolds (with or without boundary) with $\dim M = n$ and $\dim N = k$. A map $f : M \to N$ is said to be *differentiable of class $C^r$*, iff for every $p \in M$ there exist a chart $(U, \varphi)$ of $M$ around $p$ and a chart $(V, \psi)$ of $N$ around $f(p)$ with $f(U) \subseteq V$, such that $\psi \circ f \circ \varphi^{-1} : \varphi(U) \to \psi(V)$ is differentiable of class $C^r$. ∎

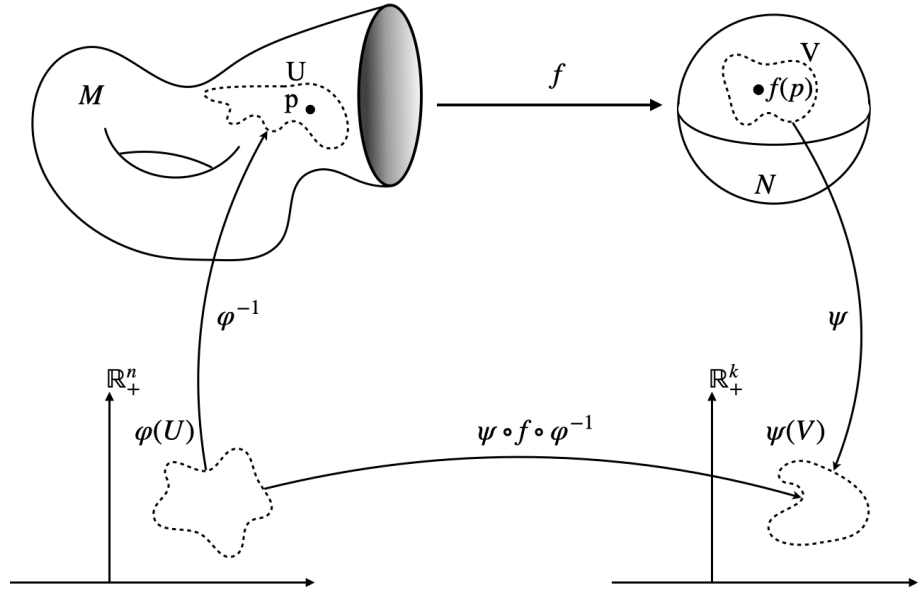

**Fig. 12**. Differentiability of a Map Between Differentiable Manifolds

Recall that $\mathbb{R}^k$ is a manifold with the atlas $\mathfrak{A} = \{(\mathbb{R}^k, \text{id})\}$. Thus, the definition before covers the definition of differentiability of maps $f : M \to \mathbb{R}^k$.

Maps that maintain the differentiable structure of manifolds are of special interest:

**Definition 7:** Let $M$ and $N$ be two $C^r$-manifolds. A map $f : M \to N$ is said to be a $C^r$-*diffeomorphism* iff $f$ is differentiable of class $C^r$, is bijective, and has an inverse that is also differentiable of class $C^r$. Furthermore, the manifolds $M$ and $N$ are called $C^r$-*diffeomorphic*. ∎

Diffeomorphic manifolds can not be distinguished based on differential topological properties. A diffeomorphism in differential topology plays the same role like homeomorphisms in general topology, or isomorphisms in algebra.

Next, we show that a chart is a diffeomorphism:

**Note 6**: Let $M$ be $C^r$-manifold (with or without boundary) with atlas $\mathfrak{A}$ and let $(U, \chi) \in \mathfrak{A}$ be a chart. Then, $\chi : U \to \chi(U)$ is a $C^r$-diffeomorphism.

Proof: First, we have to show that $\chi$ is of class $C^r$. I.e. we have to show that $\psi \circ \chi \circ \varphi^{-1} : \varphi(W) \to \psi(V)$ is of class $C^r$ for a chart $(W, \varphi)$ of $M$ and a chart $(V, \psi)$ of $\mathbb{R}^k$. Choose $(W, \varphi) = (U, \chi)$ as chart of $M$ and $(V, \psi) = (\mathbb{R}^k, \text{id})$ as chart of $\mathbb{R}^k$. Then $\psi \circ \chi \circ \varphi^{-1} = \text{id} \circ \chi \circ \chi^{-1} = \text{id} : \chi(U) \to \mathbb{R}^k$, which is of class $C^r$.

$\chi$ is a homeomorphism, i.e. $\chi$ is bijective. Similar as before it is seen that $\chi^{-1}$ is of class $C^r$. ∎

The general question is about the relevance of the differentiability class $C^r$. A famous theorem by Whitney [W36] proves that studying $C^\infty$-manifolds suffice (a detailed proof of this theorem can be found in [H76], 2-2.10). Because of this theorem the restriction to smooth manifolds (i.e. $C^\infty$-manifolds) is justified:

**Theorem 1**: Let $1 \leq r \leq \infty$. Then: Every $C^r$-manifold is $C^r$-diffeomorphic to a $C^\infty$-manifold. ∎

### 4.5. Differential of a Map

A point $p \in M$ of a differentiable manifold $M$ (with or without boundary) is associated with its *tangent space* $T_pM$ (see Figure 13). This tangent space can be imagined as the set of all tangent vectors to $M$ through $p$; the precise definition is much more subtle and complex (see [L13], chapter 3) but for our purpose this descriptive idea suffice.

$T_pM$ is a vector space, and if $M$ has dimension $m$ it is $\dim T_pM = m$ ([L13] proposition 3.12). The disjoint union of the tangent spaces of all points $p \in M$ is referred to as *tangent bundle* $TM$ of M: $TM = \biguplus_{p \in M} T_pM$. With $\dim M = m$, $TM$ is a differentiable manifold with $\dim TM = 2m$ ([L13] proposition 3.18).

If $f : M \to N$ is a differentiable map between two differentiable manifolds, then the *differential* $df_p$ of $f$ at $p \in M$ is a linear map $df_p : T_pM \to T_{f(p)}N$ (see Figure 13). As before, the precise definition is quite complex ([L13], chapter 3) but, again, a vague intuition suffice for our purpose, especially because the differential $df_p$ corresponds to the Jacobian matrix of $f$ ([L13], p. 61 ff). We will need the latter representation of the differential of a map because it allows us to compute the rank of the differential as the rank of the Jacobian matrix.

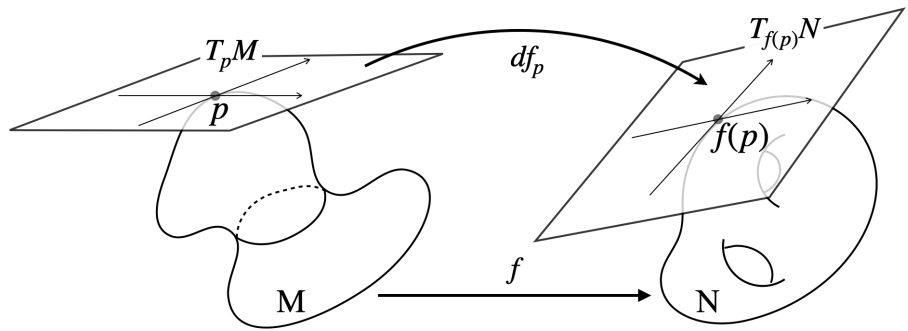

**Fig. 13**. Tangents Space and Differential

In case $M = \mathbb{R}^m$ and $N = \mathbb{R}^n$ are "just" Euclidian spaces, $M$ and $N$ are smooth manifolds (see Note 5 (a)), and their tangent spaces are the very same Euclidian spaces, i.e. $T_x\mathbb{R}^m = \mathbb{R}^m$ and $T_y\mathbb{R}^n = \mathbb{R}^n$ for any points $x \in \mathbb{R}^m$ and $y \in \mathbb{R}^n$ ([L13] proposition 3.13). The differential $df_p : T_pM \to T_{f(p)}N$ of a map $f : M \to N$ becomes the total derivative $Df(p) = \left(\partial f_i(p)/\partial x_j\right)$ ([L13] proposition C.3); thus, for $v \in T_pM$ it is $df_p(v) = Df(p)(v) = D_vf(p)$ where $D_vf(p)$ is the directional derivative of $f$ in the direction $v$.

### 4.6. Maps of Constant Rank

For each linear map $L : V \to W$ between vector spaces $V$ and $W$ the rank of $L$ is defined as the dimension of the image of $L$, i.e. $\operatorname{rank} L := \dim \operatorname{img} L$. Since the image of $L$ is a subspace of $W$, it is always $\operatorname{rank} L \leq \dim W$. According to the Fundamental Theorem of Linear Algebra it is $\operatorname{rank} L + \dim \ker L = \dim V$, thus, $\operatorname{rank} L = \dim V - \dim \ker L \leq \dim V$. This proves the following

**Note 7:** Let $L : V \to W$ be a linear map. Then: $\operatorname{rank} L \leq \min\{\dim V, \dim W\}$. ∎

The rank of differentiable maps $f : M \to N$ in $p \in M$ is the rank of the linear map $df_p$. It is key to the study of local and global properties of differentiable functions $f$.

**Definition 8:** Let $f : M \to N$ be a differentiable map between the two differentiable manifolds M and N. The *rank* of $f$ at $p \in M$ is the rank of its differential $df_p : T_pM \to T_{f(p)}N$, in symbols $\operatorname{rank}_p f$. If $f$ has the same rank $r$ at every point $p \in M$ then $f$ is said to have *constant rank*, in symbols $\operatorname{rank} f = r$. ∎

Since $df_p : T_pM \to T_{f(p)}N$ is linear, it is $\operatorname{rank}_p f \leq \min\{\dim M, \dim N\}$ (see Note 7 before). In case $\operatorname{rank}_p f = \min\{\dim M, \dim N\}$, $f$ is said to have *full rank* at $p$ and just full rank if it has full rank at every point of $M$. Maps of full rank have special names:

**Definition 9:** Let $f : M \to N$ be a differentiable map between the two differentiable manifolds M and N. $f$ is called an *immersion* in case $\operatorname{rank} f = \dim M$, i.e. $df_p$ is injective. $f$ is called a *submersion* in case $\operatorname{rank} f = \dim N$, i.e. $df_p$ is surjective. ∎

An often used property is that the rank of a map can locally not decrease:

**Lemma 5:** Let $f : M \to N$, $p \in M$ and $\operatorname{rank}_p f = k$. Then their is a neighborhood $U \in \mathfrak{U}_p$ such that $\operatorname{rank}_x f \geq k$ for each $x \in U$.

<u>Proof</u>: It is $\operatorname{rank} df_p = k$; thus, there is $k \times k$-submatrix $A(p)$ of $df_p$ with $\det A(p) \neq 0$. W.l.o.g. this submatrix is

$$A(p) = \left(\frac{\partial f_i}{\partial x_j}(p)\right)_{1 \leq i,j \leq k}$$

Next, define the map

$$\Delta : M \to \mathbb{R}, x \mapsto \det\left(\frac{\partial f_i}{\partial x_j}(x)\right)_{1 \leq i,j \leq k}.$$

According to the Leibniz formula from linear algebra it is

$$\Delta(x) = \det df_x = \sum_{\sigma \in S_k} \left( \text{sgn}(\sigma) \prod_{i=1}^{k} \frac{\partial f_i}{\partial x_{\sigma(i)}}(x) \right),$$

i.e. $\Delta$ is continuous (even differentiable) because $f$ is differentiable and the Leibniz formula is a polynomial. Thus, with $\Delta(p) = \det A(p) \neq 0$ there is an $U \in \mathfrak{U}_p$ such that $\Delta(x) \neq 0$ for each $x \in U$. Consequently, $\text{rank}_x f \geq k$ for each $x \in U$. ∎

Although the rank of a map can locally not decrease, it may increase: For example, for $f(x, y) := (y, x^2 + y)$ it is $df_{(x,y)} = \begin{pmatrix} 0 & 2x \\ 1 & 1 \end{pmatrix}$. Thus, $df_{(0,0)} = \begin{pmatrix} 0 & 0 \\ 1 & 1 \end{pmatrix}$ and $\text{rank}_{(0,0)} f = 1$. Arbitrary close to (0,0) it is $df_{(\varepsilon,\varepsilon)} = \begin{pmatrix} 0 & 2\varepsilon \\ 1 & 1 \end{pmatrix}$, i.e. for $\varepsilon > 0$ it is $\text{rank}_{(\varepsilon,\varepsilon)} f = 2$. However, if the rank is already maximal, it is locally constant since it cannot decrease locally; this proves:

**Lemma 6:** Let $f : M \to N$ be of full rank at $p \in M$. Then it is of full rank in a neighborhood $U \in \mathfrak{U}_p$. ∎

This has an important implication:

**Corollary 1:** Let $M, N$ be two differentiable manifolds and $f : M \to N$ a differentiable map. Then:
  a. If $df_p$ is surjective, then there exists a $U \in \mathfrak{U}_p$ such that $f|_U$ is a submersion.
  b. If $df_p$ is injective, then there exists a $U \in \mathfrak{U}_p$ such that $f|_U$ is an immersion.

Proof: (a) By assumption $df_p$ is surjective, i.e. by definition $\text{rank}_p f = \dim N$. But $\text{rank}_p f \leq \min\{\dim M, \dim N\}$ implies that $\dim N \leq \min\{\dim M, \dim N\}$. Thus, $\dim N = \min\{\dim M, \dim N\}$ because $\dim N < \min\{\dim M, \dim N\}$ would be a contradiction. Consequently, $\text{rank}_p f = \min\{\dim M, \dim N\}$ which shows that $f$ is of full rank. The lemma before proves the claim. (b) is proven the same way. ∎

The next theorem (whose proof can be found in [L13], Theorem 4.5) shows that in case the differential is bijective at a point of a manifold without boundary the map is a local diffeomorphism around that point. Note, that the precondition that the manifold must have no boundary is essential here: The inclusion $\iota : \mathbb{R}_+^n \hookrightarrow \mathbb{R}^n$ is corresponding counter-example.

**Theorem 2 (Inverse Function Theorem)**: Let $M, N$ be two differentiable manifolds without boundary and $f : M \to N$ a differentiable map. If $df_p$ is bijective there are $U \in \mathfrak{U}_p$ and $V \in \mathfrak{U}_{f(p)}$ such that $f|_U : U \to V$ is a diffeomorphism. ∎

In the theorem before the conditions that both manifolds must have no boundary can be weakened: the codomain $N$ may have a boundary but the image of $f$ must be in the interior of the codomain, i.e. $f(M) \subseteq N \setminus \partial N$; the reason is that $N \setminus \partial N \subseteq_{\text{open}} N$ is a submanifold without boundary (see Note 5 (c)), i.e. the original inverse function theorem applies.

Special kinds of immersions play an important role:

**Definition 10:** Let $f : M \to N$ be a differentiable map between the two differentiable manifolds M and N. $f$ is called an *embedding* iff $f$ is an immersion and $f : M \to f(M)$ is a homeomorphism onto $f(M) \subseteq N$ in the subspace topology. ∎

The map $f : \mathbb{R} \to \mathbb{R}^2$, $x \mapsto (x^3, 0)$ is a homeomorphism, $f$ is smooth, but because of $df_0 = 0$, $f$ is not an immersion, thus, no embedding. Thus, not every smooth homeomorphism is automatically an embedding. However, under the following condition injective immersions are already embeddings (the proof can be found in [L13], Proposition 4.22):

**Lemma 7:** Let $f : M \to N$ be an injective immersion, $M$ and $N$ be manifolds with or without boundary. If any of the following holds, $f$ is an embedding:
  a. M is compact.
  b. M has no boundary, and dim $M$ = dim $N$. ∎

The following theorem is the basis of many other theorems in differential topology (see [AA22], Theorem 3.7.5):

**Theorem 3 (Rank Theorem)**: Let $M, N$ be two differentiable manifolds and $f : M \to N$ a differentiable map. Let $p \in M$, $q := f(p) \in N$ and let $U \in \mathfrak{U}_p$ be a neighborhood of $p$ such that for each $x \in U$ it is $\text{rank}_x f = r$ (i.e. $f$ has constant rank in $U$). Then there is a chart for $M$ around $p$ and a chart for $N$ around $q$, such that $f$ has in these charts the form $f(x_1, \ldots, x_r, x_{r+1}, \ldots, x_m) = (x_1, \ldots, x_r, 0, \ldots, 0)$. ∎

Several key properties of maps are inherited by their composition:

**Note 8:** Let $L, M, N$ be differentiable manifolds, and let $f : L \to M$ and $g : M \to N$ be maps. Then:
  a. If $f$ and $g$ are injective or surjective or bijective then $g \circ f$ is injective or surjective or bijective.
  b. If $f$ and $g$ are immersions then $g \circ f$ is an immersion.
  c. If $f$ and $g$ are continuous then $g \circ f$ is continuous.
  d. If $f$ and $g$ are homeomorphisms then $g \circ f$ is a homeomorphism.
  e. If $f$ and $g$ are embeddings then $g \circ f$ is an embedding.

<u>Proof</u>: (a) $g(f(x)) = g(f(y))$ implies $f(x) = f(y)$ because $g$ is injective. Next, injectivity of $f$ implies $x = y$. This proves the injectivity of $g \circ f$.

$f(L) = M$ and $g(M) = N$, thus, $(g \circ f)(L) = N$. This proves the surjectivity of $g \circ f$. Together, this proves the bijectivity of $g \circ f$.

(b) $df$ and $dg$ are injective. According to the chain rule it is $d(g \circ f) = dg \circ df$, i.e. part (a) shows that $d(g \circ f)$ is injective. Thus, $g \circ f$ is an immersion,

(c) Let $O \subseteq N$ be open. Because $g$ is continuous, $g^{-1}(O) \subseteq M$ is open. Because $f$ is continuous, $f^{-1}(g^{-1}(O)) \subseteq L$ is open. With $f^{-1} \circ g^{-1} = (g \circ f)^{-1}$ it follows that $(g \circ f)^{-1}(O) \subseteq L$ is open. Thus, $g \circ f$ is continuous.

(d) $f$ and $g$ are bijective, so is $g \circ f$ (see part (a)). $f$ and $g$ are continuous, so is $g \circ f$ (see part (c)). $f^{-1}$ and $g^{-1}$ are continuous, so is $(g \circ f)^{-1} = f^{-1} \circ g^{-1}$ (see part (c)). Thus, $g \circ f$ is a homeomorphism.

(e) $f$ and $g$ are embeddings, i.e. both are immersions as well as homeomorphisms. Because of part (b) $g \circ f$ is an immersion, and because of part (d) $g \circ f$ is a homeomorphism. Thus, $g \circ f$ is an embedding. ∎

### 4.7. Submanifolds

It is important to note that the definition of a manifold is completely independent of any surrounding space like a Euclidian space. In this sense, manifolds are abstract entities. Their concept has been introduced by Bernhard Riemann in 1854 (published 1868 [R]). It generalizes objects like curves and surfaces that had been studied before that time. The latter are entities within a Euclidian space. Thus, it is natural to ask whether any (abstract) manifold is "equivalent" (a.k.a. "diffeomorphic") to a corresponding entity in a Euclidian space: this has been proven by Hassler Whitney in 1936 [W36]. In this section, we summarize the corresponding concepts and results as relevant in our context.

A manifold maybe contained in another manifold (see Figure 14).

**Definition 11:** Let $M$ be differentiable $C^r$-manifold, $\dim M = n$ and let $S \subset M$ be a subset of $M$. $S$ is called an *embedded submanifold* of $M$ *of dimension $k$* (or of *codimension $n - k$*, respectively) and class $C^r$ if for any point $p \in S$ there is a chart $(U, \varphi)$ of $M$ around $p$ such that $\varphi(U \cap S) = \varphi(U) \cap (\mathbb{R}^k \times \{0\})$. ∎

Sometimes, embedded submanifolds are also called *regular submanifolds*. Figure 14 depicts the situation. If $\mathfrak{A} = \{(U_i, \varphi_i) | i \in I\}$ is an atlas of $M$ then $\mathfrak{A}|_S := \{(U_i \cap S, \varphi_i|_{U_i \cap S}) | i \in I\}$ is an atlas of $S$. Note that the latter assumes that $S$ is a topological subspace of $M$, i.e. the topology of $S$ is the subspace topology. This will become important soon.

Especially, since $\mathfrak{A}|_S$ is an atlas of $S$, $S$ is a $C^r$-manifold in itself:

**Note 9:** Every embedded submanifold $S$ of class $C^r$ and dimension $k$ is a $C^r$-manifold with $\dim S = k$. ∎

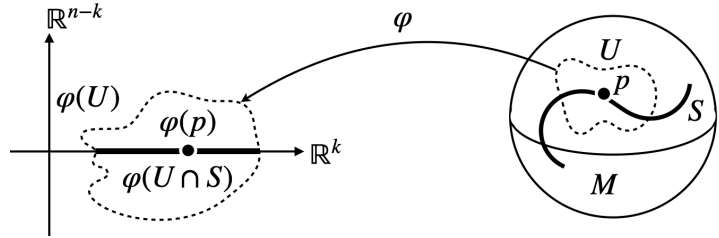

**Fig. 14**. A Chart of an Embedded Submanifold

The following lemma motivates the name "embedded" submanifold (see [T], Theorem 11.14 for a proof):

**Lemma 8**: Let $M$ be differentiable $C^r$-manifold, and let $S \subset M$ be an embedded submanifold of $M$. Then, the inclusion $\iota : S \hookrightarrow M$ is a $C^r$-embedding (and, thus, by definition an immersion). ∎

Vice versa, the name "embedding" of a map is justified by the following lemma (see [T], Theorem 11.13 for a proof):

**Lemma 9**: Let $M$ and $S$ be differentiable $C^r$-manifolds, and let $f : S \to M$ be a $C^r$-embedding. Then, the image $f(S) \subseteq M$ is an embedded submanifold of $M$. ∎

A two-dimensional surface is an example of a manifold embedded in the three-dimensional Euclidian space. In general, graphs of differentiable functions are examples of such surfaces. The following lemma and definition generalizes the corresponding situation (see [L13], Proposition 5.4):

**Lemma & Definition 10**: Let $M$ and $N$ be $C^r$-manifolds ($N$ with or without boundary), $\dim M = m$ and $\dim N = n$, $U \subseteq_{\text{open}} M$ and $f : U \to N$ be of class $C^r$. Then the graph $\Gamma(f) \subseteq M \times N$ is an embedded $C^r$-manifold of dimension $m$ without boundary.
Here, the *graph of f* is defined as $\Gamma(f) := \{(x, y) \in M \times N \,|\, x \in U \wedge y = f(x)\}$. ∎

The Euclidian space $\mathbb{R}^n$ is a manifold, so its embedded submanifolds are of interest. The following is an often used mechanism to produce embedded submanifolds in $\mathbb{R}^n$: For $U \subseteq_{\text{open}} \mathbb{R}^k$ and a $C^r$-differentiable map $f : U \to \mathbb{R}^m$ the graph $\Gamma(f) \subseteq \mathbb{R}^k \times \mathbb{R}^m$ is an embedded $C^r$-manifold of dimension $k$ without boundary (according to the lemma just before). If $f : U \subseteq \mathbb{R}^{n-1} \to \mathbb{R}$ is of class $C^r$, the graph $\Gamma(f)$ is a *hypersurface* in $\mathbb{R}^n$, i.e. a $C^r$-manifold of dimension $n-1$ (see Figure 15); obviously, this generalizes the notion of a surface in $\mathbb{R}^3$.

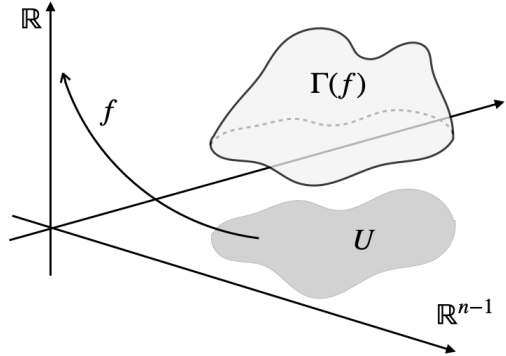

**Fig. 15**. The Graph of a Map as Manifold

Another important means to get embedded submanifolds is via so-called level-sets and regular values:

**Definition 13:** Let $f : M \to N$ be a $C^r$-map.
  a. For any point $q \in N$ the set $f^{-1}(q)$ is called a *level-set of f*.
  b. A point $p \in M$ is called a *regular point* if $df_p$ is surjective ($p$ is called a *critical point* otherwise).
  c. $q \in N$ is called a *regular value* if each point of $f^{-1}(q)$ is a regular point ($q$ is called *critical value* otherwise).
  d. If $q \in N$ is a regular value the level-set $f^{-1}(q)$ is called a *regular level-set*. ∎

Obviously, each point of $M$ is a critical point if $\dim M < \dim N$ ($df_p$ cannot be surjective in this case). Any regular level set is a $C^r$-manifold (see [L13], Corollary 5.14):

**Lemma 11**: Let $M$ and $N$ be $C^r$-manifolds, and $f : M \to N$ be of class $C^r$. Then, any regular level-set $f^{-1}(q) \subseteq M$ is an embedded submanifold of $M$ with $\dim f^{-1}(q) = \dim M - \dim N$. ∎

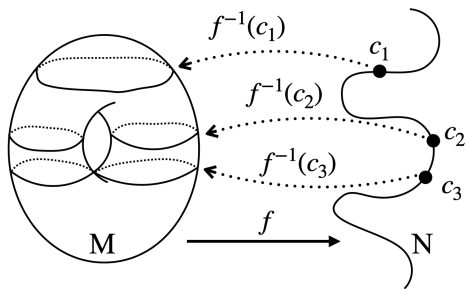

**Fig. 16**. Level-Sets of a Differentiable Map

In Figure 16, $f^{-1}(c_1), f^{-1}(c_2)$, and $f^{-1}(c_3)$ are three level-sets of the differentiable map $f$. Assuming that both, $c_1$ and $c_2$ are regular values, the two level sets $f^{-1}(c_1)$ and $f^{-1}(c_2)$ are embedded submanifolds (the first one diffeomorphic to a circle, the second one diffeomorphic to two disjoint circles) of $M$ of dimension 1 because of $\dim M = 2$ and $\dim N = 1$. But $f^{-1}(c_3)$ is not an embedded submanifold because it has the shape of an "8", i.e. it contains a singularity in the form of a self-intersection (see section 4.3). This implies that $c_3$ is a critical value.

The lemma before allows to prove that the set of unitary transformations of a complex vector space is a manifold:

**Lemma 12:** $U(n)$ is a smooth compact connected manifold (even an embedded submanifold) with $\dim_{\mathbb{R}} U(n) = n^2$. $U(n)$ is even a Lie-group.

Proof: It is $U(n) = \{A \in \mathrm{GL}(n, \mathbb{C}) \mid A^*A = I\}$, and $\mathrm{GL}(n, \mathbb{C}) = \mathbb{C}^{n^2} = \mathbb{R}^{2n^2}$; thus, it is $U(n) \subset \mathbb{R}^{2n^2}$. Define $f : \mathbb{R}^{2n^2} = \mathrm{GL}(n, \mathbb{C}) \to \mathrm{GL}(n, \mathbb{C}) = \mathbb{R}^{2n^2}$ by $A \mapsto A^*A$; $f$ is differentiable because building the conjugate transpose of a matrix is differentiable and the multiplication of two matrices is differentiable too. The differential of $f$ is $df_A(V) = A^*V + V^*A$ (see Appendix A).

Let $H(n) = \{A \in \mathrm{GL}(n, \mathbb{C}) \mid A = A^*\}$ be the set of all Hermitian matrices. $H(n)$ is a $\mathbb{R}$-vector space of dimension $n^2$ ([LM15] Lemma 13.15). Thus, $H(n)$ is a smooth manifold of dimension $n^2$.

Because $f(B)^* = (B^*B)^* = B^*(B^*)^* = B^*B = f(B)$ for each $B \in \mathrm{GL}(n, \mathbb{C})$, it is $f(\mathrm{GL}(n, \mathbb{C})) \subseteq H(n)$, i.e. $f$ is in fact a map $f : \mathrm{GL}(n, \mathbb{C}) \to H(n)$ (and $f$ is differentiable).

Next we show that any unitary map $A \in U(n)$ is a regular point of $f : \mathrm{GL}(n, \mathbb{C}) \to H(n)$, i.e. that the differential $df_A$ is surjective for each $A \in U(n)$:

Choose an arbitrary $B \in H(n)$ and define $W := \frac{1}{2}AB \in \mathrm{GL}(n, \mathbb{C})$. Then,
$$df_A(W) = \frac{1}{2}A^*AB + \frac{1}{2}B^*A^*A \stackrel{(1)}{=} \frac{1}{2}IB + \frac{1}{2}B^*I = \frac{1}{2}B + \frac{1}{2}B^* \stackrel{(2)}{=} \frac{1}{2}B + \frac{1}{2}B = B$$
(hereby, (1) is because $A^*A = I$ and (2) because $B^* = B$).

Now it is $f^{-1}(I) = U(n)$, thus $I \in H(n)$ is a regular value and $U(n) = f^{-1}(I)$ is a regular level set. Consequently, $U(n)$ is an embedded submanifold (Lemma 11), and, thus, a manifold itself (Note 9); furthermore, $\dim U(n) = \dim f^{-1}(I) = \dim \mathrm{GL}(n, \mathbb{C}) - \dim H(n) = 2n^2 - n^2 = n^2$.

Next, we prove compactness.

(a) $f : \mathrm{GL}(n, \mathbb{C}) \to \mathrm{GL}(n, \mathbb{C})$ with $A \mapsto A^*A$ is especially continuous. Now, $U(n) = f^{-1}(I)$, $\{I\} \subset \mathrm{GL}(n, \mathbb{C})$ is a closed set and pre-images of closed sets under continuous maps are closed, i.e. $U(n) \subset_{\text{closed}} \mathbb{R}^{2n^2}$.

(b) Next, $A^*A = I$ means especially that the columns of $A = (a_{ij})$ are unit vectors, i.e. $\sum_i |a_{ij}|^2 = 1$ for each $1 \leq j \leq n$. Thus, $|a_{ij}|^2 \leq 1$ for all $1 \leq i, j \leq n$. This implies that $\|A\|^2 = \sum_{i,j=1}^n |a_{ij}|^2 \leq n^2$ for each $A \in U(n)$, i.e. $U(n)$ is bounded in $\mathbb{R}^{2n^2}$. Thus, according to the theorem of Heine-Borel, $U(n)$ is compact in $\mathbb{R}^{2n^2} = \mathrm{GL}(n, \mathbb{C})$.

Finally, we prove connectedness.

Le $A \in U(n)$. Then, $A$ is diagonalizable, i.e. there exists a unitary matrix $T$ such that $A = T \text{diag}\left(e^{i\varphi_1}, \cdots, e^{i\varphi_n}\right) T^*$ ([LM15] Theorem 18.13). Let $0 \leq t \leq 1$; then, $A_t := T \text{diag}\left(e^{it\varphi_1}, \cdots, e^{it\varphi_n}\right) T^* \in U(n)$ (because $A, B \in U(n) \Rightarrow AB \in U(n)$). It is $A_0 = I$ and $A_1 = A$, i.e. there is a path from the identity matrix $I$ to $A$. Thus, any two unitary matrices can be connected by a path (e.g. via $I$): $U(n)$ is path-connected. Since every path-connected topological space is connected ([P22] Proposition 9.26), $U(n)$ is connected.

$U(n)$ is a Lie group because matrix multiplication is differentiable map $\cdot : U(n) \times U(n) \to U(n)$, $(a_{ij}) \cdot (b_{st}) = \left(\sum_j a_{ij} b_{jt}\right)_{it}$, and for two unitary matrices $A, B \in U(n)$ it is $(AB)(AB)^* = ABB^*A^* = AIA^* = AA^* = I$, i.e. $AB \in U(n)$. ∎

Note, that the condition "regular" is key in Lemma 11. Without that condition, any closed subset of a manifold can be made the level set of a differentiable function (see [L13], Theorem 2.29 for a proof):

**Lemma 13**: Let $M$ be a differentiable $C^r$-manifold, and let $K \subseteq_{\text{closed}} M$ be a closed subset of $M$. Then there exists a $C^r$-function $f : M \to \mathbb{R}$ with $f^{-1}(0) = K$. ∎

In many practical situation, both, $M$ and $N$ are Euclidian spaces. For example, with $M = \mathbb{R}^{n+1}$ and $N = \mathbb{R}$, the map $f : \mathbb{R}^{n+1} \to \mathbb{R}$, $x \mapsto |x|^2$ has the differential $df_x = 2(x_1 \cdots x_n) \neq 0$ for $x \neq 0$, i.e. $\text{rank}_x f = 1$ for $x \neq 0$, thus, $df_x$ is surjective except at the origin. This implies that each $c \neq 0 \in \mathbb{R}$ is a regular value; according to the lemma before, $f^{-1}(c)$ is a regular level-set for each $c \neq 0$. Thus, $f^{-1}(c) = \{x \in \mathbb{R}^{n+1} \mid |x|^2 = c\}$ is a sphere of dimension $n$, especially the unit sphere $\mathbb{S}^n = f^{-1}(1)$ is an $n$-dimensional embedded submanifold of $\mathbb{R}^{n+1}$.

There are important situations (e.g. in the context of Lie groups - see [S21]) in which the notion of a submanifold has to be generalized. Very roughly, any differentiable manifold that is a subset of another differentiable manifold and is "properly situated" there is considered a certain kind of a submanifold. More precise:

**Definition 14:** Let $M$ and $N$ be $C^r$-manifolds and $f : N \to M$ be an injective immersion of class $C^r$. Then, the image $f(N) \subseteq M$ is called an *immersed submanifold* of $M$. ∎

For example, let $f : ]-\pi, +\pi[ \to \mathbb{R}^2$, $x \mapsto (\sin 2x, \sin x)$. Then, $f$ is injective on the open interval $]-\pi, +\pi[$ (see Appendix B for more details). Also, $f$ is an immersion because $df_x = (2\cos 2x, \cos x) \neq 0$, i.e. $\text{rank}_x f = 1$ for $x \in ]-\pi, +\pi[$. Thus, $f$ is an injective $C^r$-immersion, i.e. $S := f(]-\pi, +\pi[) \subseteq \mathbb{R}^2$ is an immersed submanifold (see Figure 17). Furthermore, $f : ]-\pi, +\pi[ \to S$ is also surjective, thus, bijective, but $f$ is not a homeomorphism because the image $S \subseteq_{\text{compact}} \mathbb{R}^2$ is compact with the subspace topology (it is closed and bounded), $]-\pi, +\pi[ \subseteq \mathbb{R}$ is not compact, while compactness is a topological invariant. Together, $S \subseteq \mathbb{R}^2$ is an immersed submanifold but $S$ is not an embedded submanifold of the manifold $\mathbb{R}^2$.

Another argument supporting the latter: any neighborhood of the origin of $S$ contains a singularity, namely a shape like in Figure 11 (b).

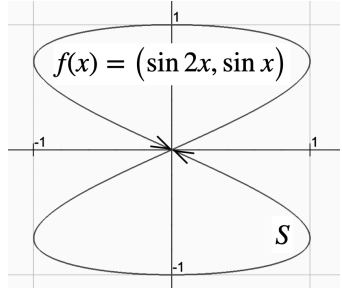

**Fig. 17**. An Injective Immersion That is No Topological Embedding

According to Lemma 8, for any embedded submanifold $S \subset M$ the inclusion $\iota : S \hookrightarrow M$ is an embedding and, thus, an injective immersion, i.e. $\iota(S) = S \subset M$ is an immersed submanifold of $M$. This proves:

**Lemma 14**: Let $M$ be a $C^r$-manifold (with our without a boundary), $S \subset M$ an embedded submanifold. Then, $S$ is an immersed submanifold. ∎

The opposite is not true: The example before shows that an immersed submanifold is in general not an embedded submanifold. However, the following lemma ([L13], Proposition 5.21) gives two situations in which an immersed submanifold is already an embedded submanifold:

**Lemma 15**: Let $M$ be a $C^r$-manifold (with our without a boundary), $S \subseteq M$ an immersed submanifold. If $\dim S = \dim M$ or if $S \subseteq_{\text{compact}} M$ then $S$ is embedded. ∎

Finally, we answer the question posed at the beginning of this section: Any compact (abstract) manifold is diffeomorphic to an embedded submanifold of a Euclidian space of high enough dimension - more precise:

**Theorem 4**: Let $r \geq 1$ and let M be a compact $C^r$-manifold with boundary. M can be embedded into $\mathbb{R}_+^{2m+1}$ with $\partial M \subseteq \partial \mathbb{R}_+^{2m+1} = \left\{ x \in \mathbb{R}^{2m+1} \mid x_{2m+1} = 0 \right\}$. ∎

Again, this theorem is by Whitney [W36], and a detailed proof can be found in [H76] (Theorem 1-4.3).

### 4.8. Volumes of Manifolds and the "Uniform Approach"

Section 2.3.1 motivated to define the expressivity of a variational quantum circuit by the "volume" of $\Omega_{\mathcal{C}}(\mathbb{P})$ in the unitary group $U(n)$: vividly, the larger this volume the higher the likelihood that $\Omega_{\mathcal{C}}(\mathbb{P})$ hits the solutions $\mathcal{S}(X)$ of a given problem $X$.

*4.8.1. Linear Approximations*

Here, we provide more details about the notion of "the volume" of a differentiable manifolds and especially how the notion of "volume" is related to the unitary group.

For this purpose, we first remind that a function *f* is called differentiable at a point *x* if it can be approximated locally by a linear function:

$$f(x + \xi) = f(x) + L\xi + o(\|\xi\|) \tag{6}$$

Here, $\xi$ is a point in a small neighborhood of *x*, *L* is a linear function, and $o(\|\xi\|)$ is the small error made when considering $f(x) + L\xi$ as the value of $f(x + \xi)$: in a small neighborhood the differentiable function *f* is nearly the linear function *L*. For a differentiable function $f : \mathbb{R} \to \mathbb{R}$ this means that *L* is a real number and $f(x) + L\xi$ is the tangent at *x* at the graph of *f*; this tangent locally approximates the function *f*. Thus, the graph of *f*, i.e. the manifold $\Gamma(f)$ (see Lemma 10), is approximated by this tangent around $(x, f(x))$.

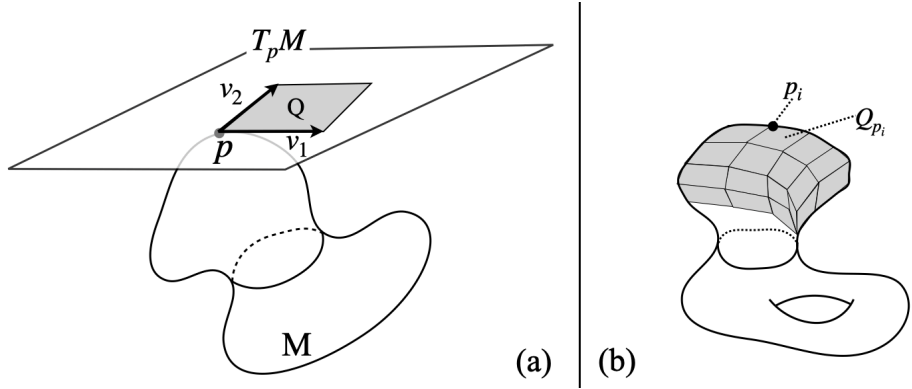

**Fig. 18**. Linear Approximation of a Differentiable Manifold

The idea of linear approximation can be used in arbitrary dimensions: in part (a) of Figure 18, a basis $v_1, v_2$ has been chosen for the tangent space $T_pM$ of the manifold *M* at point *p*. The basis spans a parallelepiped $Q := \{a_1v_1 + a_2v_2 \mid 1 \leq a_1, a_2 \leq 1\}$. If we take small vectors $v_1, v_2$ the parallelepiped approximates the manifold *M* around *p* with a small error, i.e. the manifold looks locally like a very small parallelepiped of the tangent space.

Applying this linear approximation for enough points $p_1, \ldots, p_k$ of the manifold, i.e. if the manifold is "covered" by parallelepipeds, the manifold is turned into a linear approximation of the whole manifold: part (b) of Figure 18 depicts this for the upper part of a manifold which looks like small parallelepipeds glued together (in the direction of the tangent spaces).

*4.8.2. Approximating Volumes*

Such a linear approximation allows to compute the approximate volume of the manifold vol(*M*) by computing the volume of the parallelepipeds vol($Q_{p_i}$) and summing up their volumes:

$$\text{vol}(M) \approx \sum_{i=1}^{k} \text{vol}(Q_{p_i}) \tag{7}$$

The volume of a parallelepiped is computed as follows: Let $v_1, \ldots, v_n$ be linear independent vectors, and let $Q := \{\sum_i a_i v_i \mid 1 \leq a_1, a_2, \ldots, a_n \leq 1\}$ be the parallelepiped spanned by these vectors. Then, from linear algebra it is known that the volume $\text{vol}(Q)$ of the parallelepiped is

$$\text{vol}(Q) = \left| \det(v_1 v_2 \ldots v_n) \right| \tag{8}$$

Define $V := (v_1 v_2 \ldots v_n)$ to be the matrix with columns $v_1, v_2, \ldots, v_n$. Then:

$$\begin{aligned}
\text{vol}(Q)^2 = |\det(V)|^2 &= \det(V) \cdot \det(V) \\
&= \det(V^T) \cdot \det(V) \\
&= \det(V^T \cdot V) \\
&= \det\left(\langle v_i, v_j \rangle\right)_{1 \leq i,j \leq n}
\end{aligned}$$

Thus,

$$\text{vol}(Q) = \sqrt{\det\left(\langle v_i, v_j \rangle\right)_{1 \leq i,j \leq n}} \tag{9}$$

*4.8.3. Riemannian Manifolds and Volume Forms*

Equation (9) reveals that computing the volume of a parallelepiped depends on a scalar product. Consequently, we need a scalar product for every tangent space of the manifold $M$.

**Definition 15:** Let $M$ be a differentiable manifold, and let $g$ be a function that associates with each $p \in M$ a scalar product $g_p : T_pM \times T_pM \to \mathbb{R}$ "in a differentiable manner". Then, $g$ is called a *Riemannian metric* on $M$, and $(M, g)$ is called a *Riemannian manifold*. ∎

Note, that the phrase "in a differentiable manner" is left vague: a precise definition would require to define differentiable vector fields which we don't need in this paper. Also, the differentiability of $g$ is not relevant in our context.

Consequently, we assume that $M$ is a Riemannian manifold; in fact, every differentiable manifold is a Riemannian manifold ([L18], Proposition 2.4). Then, the volume of a parallelepiped $Q_p$ in $T_pM$ is:

$$\text{vol}(Q_p) = \sqrt{\det\left(g_p(v_i, v_j)\right)_{1 \leq i,j \leq n}} \tag{10}$$

With equation (7) and equation (10) we can approximate the volume of a Riemannian manifold $M$ as follows:

$$\text{vol}(M) \approx \sum_{i=1}^{k} \text{vol}(Q_{p_i}) = \sum_{i=1}^{k} \sqrt{\det\left(g_p(v_i, v_j)\right)_{1 \leq i,j \leq n}} \tag{11}$$

By choosing infinitesimal small parallelepipeds and correspondingly more and more points from the manifold we perform a limit process. In analogy to the limit process that defines the Riemannian integral we write very informally (and only conceptually) with $G = \left(g_p(v_i, v_j)\right)_{i,j}$:

$$\text{vol}(M) = \int_M \sqrt{\det(G)}\, dx \tag{12}$$

The precise definitions behind this notion need a lot more concepts and machinery. The most fundamental concept needed is that of a volume form $dV$ that abstracts our informal notation $\sqrt{\det(G)}\, dx$. In our context, the unitary group admits such a volume form:

**Note 10**: The unitary group $U(n)$ admits a (unique) volume form $dV$.

Proof: Every differentiable manifold is a Riemannian manifold ([L18], Proposition 2.4). If a Riemannian manifold $M$ is oriented then it admits a (unique) volume form $dV$ ([L18], Proposition 2.41). According to Lemma 12, $U(n)$ is a differentiable manifold, thus, it is a Riemannian manifold also. Any Lie group is orientable ([K22], Lemma 7). Since $U(n)$ is a Lie group (Lemma 12) it admits a (unique) volume form. ∎

The volume form $dV$ of a manifold is (in local coordinates) $dV = \sqrt{\det(G)}\, dx$; it allows (as equation (12) indicates) to compute the volume of $M$, namely $\text{vol}(M) = \int_M dV$. It also allows to compute the integral of functions $f$, i.e. $\int_M f\, dV$ ([see L18], the discussion following Proposition 2.41).

*4.8.4. Haar Measure*

Computing volumes is tight to differentiable manifolds, not applicable to other "spaces". For this purpose, the concept of a *measure* is introduced (see [A20] for details) that is more abstract than a volume but mimics its properties. Luckily, in our context, both concepts are the same. This is roughly seen as follows:

Each open set $W$ of a differentiable manifold $M$ is a differentiable manifold by itself (Note 5). Thus, the inclusion map $\iota : W \hookrightarrow M$ induces a Riemannian metric on $W$, and $\text{vol}(W) = \int_W dV$ is defined. This in turn defines a measure on the Borel sets $\mathcal{B}(M)$ of $M$, where $\mathcal{B}(M)$ is the smallest $\sigma$-algebra containing all open sets of $M$ (refer to [A20] for details about Borel sets, measure spaces, and measures). This turns the manifold $M$ into a measure space $(M, \mathcal{B}(M), \text{vol})$ ([G24], section 1.7). A measure is more general than computing volumes via integrals.

As a Lie group, $U(n)$ admits a left-invariant measure, the so-called *Haar measure*. This measure is unique up to multiplication by a positive constant ([S21], Theorem 3.1). A compact Lie group admits a bi–invariant Riemannian metric ([AB10], Proposition 2.17) which induces a bi-invariant volume form. Since the Haar measure is unique, we get:

**Note 11:** The volume form $dV$ on a compact Lie group is the Haar measure. ∎

Especially, since $U(n)$ is a Lie group, instead of abstractly speaking of the measure of a subset of a manifold, we can speak about its volume.

*4.8.5. Volume of Submanifold*

Having the ability to determine the volume of manifolds, the question of volumes of submanifolds comes up:

**Note 12**: Let $M$ be a manifold with volume element $dV$, and let $N \subseteq M$ be an (embedded or immersed) submanifold. If $\dim N < \dim M$ then $\text{vol}(N) = 0$.

Proof: Let $M, N$ be differentiable manifolds and let $f : N \to M$ be differentiable map. If $\dim N < \dim M$ then $f(N) \subset M$ has measure zero ([L13], Corollary 6.11). For an (embedded or immersed) submanifold $N$ of $M$ the inclusion map $\iota : N \hookrightarrow M$ is differentiable (Lemma 8), i.e. if $\dim N < \dim M$ then $N$ has measure zero in $M$. ∎

This implies that any submanifold $S$ of the unitary group $U(n)$ with $\dim S < \dim U(n)$ has measure zero. Since measure and volume coincide it is $\text{vol}(S) = 0$ for $\dim S < \dim U(n)$:

**Note 13:** Let $S \subset U(n)$ be a submanifold, $\dim S < \dim U(n)$. Then $\text{vol}(S) = 0$. ∎

Thus, $\text{vol}(\Omega_{\mathscr{C}}(\mathbb{P})) = 0$ in case $\dim \Omega_{\mathscr{C}}(\mathbb{P}) < \dim U(n)$ (if $\Omega_{\mathscr{C}}(\mathbb{P})$ is a submanifold at all - see next chapter). Consequently, the illustrative motivation about measuring expressivity in the "Unitary Approach" (section 2.3.1) requires refinement.

*4.8.6. Uniform Distribution*

Let $A = \,]0,\varepsilon[\, \times \,]0,\varepsilon[\, \subset \mathbb{R}^2$ be a small open square in the Euclidian plain. If a point has to be picked randomly from $A$, the probability for any two points $x, y \in A$ being picked is the same: picking is "uniformly random". This means that the probability of a certain point being picked is identical to the probability of any other point in $A$ being picked: the corresponding distribution of probabilities is uniform.

Obviously, the probability of a point being picked from $A$ is related to the volume of $A$: the smaller $A$ the higher the likelihood of a point within $A$ to be picked. If we move the square around in the Euclidian plain, e.g. translating it by a vector $v \in \mathbb{R}^2$ to $A_v = v + A := \{v + t \,|\, t \in \,]0,\varepsilon[\, \times \,]0,\varepsilon[\,\} \subset \mathbb{R}^2$, the probability distribution remains the same because the volume is unchanged. This is based on the left-invariance of the volume in Euclidian space: by adding $v \in \mathbb{R}^2$ from left to each vector in $A$ doesn't change the volume, i.e. $\text{vol}(A) = \text{vol}(A_v) = \text{vol}(v + A)$.

This is different for arbitrary manifolds: the Euclidian space if flat, but picking points from areas on curved manifolds may behave differently. For example, let $\mathbb{S}^2$ be the unit sphere. Then

$$f : \,]0,\pi[\, \times \,]0,2\pi[\, \to \mathbb{S}^2 \qquad (13)$$
$$(\vartheta, \varphi) \mapsto (\sin \vartheta \cos \varphi, \sin \vartheta \sin \varphi, \cos \vartheta)$$

is a chart of $\mathbb{S}^2$ (leaving out both poles as well the meridian of the sphere - otherwise $f$ would not be a diffeomorphism). Figure 19 depicts the images under $f$ of different parts of the domain:

- Part (a) shows the image of the whole domain, i.e. the sphere $\mathbb{S}^2$.

- Part (b) is the image of $]\pi/2 - \varepsilon, \pi/2 + \varepsilon[ \times ]0,2\pi[$ ($\varepsilon$ is a small positive number), resulting in a belt around the equator.
- Part (c) is the image of $]\pi - \varepsilon, \pi[ \times ]0,2\pi[$ resulting in a cap of the north pole.
- Part (c) is the image of $]0,\varepsilon[ \times ]0,2\pi[$ resulting in a cap of the south pole.

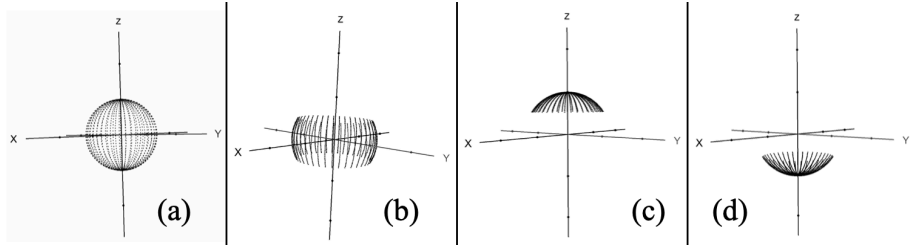

**Fig. 19**. Linear Approximation of a Differentiable Manifold

The figure indicates that the volume of the belt is larger than the volume of a cap. But a larger volume of an area means a smaller probability of a point of the area to be randomly picked. Thus, points on the sphere with values of $\vartheta$ close to $\pi/2$ (i.e. points from the belt) have a smaller probability to be randomly picked than points with values of $\vartheta$ close to $\pi$ (i.e. points from the cap of the north pole) or close to 0 (i.e. points from the cap of the south pole): by picking a random latitude, points close to the poles have a higher probability to be picked than points near the equator. In this sense, points on the sphere are somehow "concentrated" towards the poles. As a consequence, the corresponding probability distribution is not uniform.

The volume that represents the probability distribution of randomly picking a point must reflect this effect of concentration to become a uniform distribution. As the figure indicates, shifting the belt to a pole does change its volume, i.e. the "usual" volume on the sphere is not translation invariant. But the Haar measure is translation invariant by definition. Thus, if we take the Haar measure to compute the probability distribution (i.e. the volume of areas of the sphere), the distribution becomes uniform: every area of the sphere of the same Haar measure has equal probability of containing a particular chosen point.

Consequently, the difference between the Haar measure of an area and its "usual" volume is an indicator of how much a distribution based on the usual volume deviates from being uniform.

*4.8.7. Measuring Expressivity*

"Deviation" can be assessed by various means. Especially, whether "volume" is computed directly or indirectly may differ. An example of an "indirect approach" is described next and is based on [SJA19].

Whenever $A \in U(n)$ has been uniformly random chosen (often called "Haar random"), $A|0\rangle$ is a uniformly random state. Thus, if $S \subseteq U(n)$ is a set of Haar random unitary matrices then $S|0\rangle := \{A|0\rangle \mid A \in S\}$ is a Haar random set of states. But $\Omega_{\mathscr{C}}(\mathbb{P}) \subset U(n)$ is not necessary Haar random, thus, the set $\Omega_{\mathscr{C}}(\mathbb{P})|0\rangle$ is not necessarily a set of uniformly random states.

Different approaches have been defined to compare $S|0\rangle$ and $\Omega_{\mathscr{C}}(\mathbb{P})|0\rangle$. One approach (see [SJA19], [H+21]) is as follows: first, the elements of these sets are turned into matrices, i.e. for $|\psi\rangle \in S|0\rangle$ and $|\varphi\rangle \in \Omega_{\mathscr{C}}(\mathbb{P})|0\rangle$ the density matrices $|\psi\rangle\langle\psi|$ and $|\varphi\rangle\langle\varphi|$ are taken. Note, that in fact $|\varphi\rangle = |\varphi_p\rangle$ depends on parameters $p \in \mathbb{P}$. For $S = U(n)$, the matrix $\int_{U(n)} |\psi\rangle\langle\psi| \, d\mu$ is computed as well as the matrix $\int_{\mathbb{P}} |\varphi_p\rangle\langle\varphi_p| \, dx$. Finally, the deviation between these two matrices is assessed and taken as expressivity of the ansatz $\mathscr{C}$. For example, based on a matrix norm, the expressivity $\eta(\mathscr{C})$ becomes

$$\eta(\mathscr{C}) := \left\| \int_{U(n)} |\psi\rangle\langle\psi| \, d\mu - \int_{\mathbb{P}} |\varphi_p\rangle\langle\varphi_p| \, dx \right\| \tag{14}$$

The smaller $\eta(\mathscr{C})$, the closer $\Omega_{\mathscr{C}}(\mathbb{P}) \subset U(n)$ becomes to be Haar random. If $\mathscr{C}$ and $\hat{\mathscr{C}}$ are two ansatzes, ansatz $\mathscr{C}$ is called more expressive than ansatz $\hat{\mathscr{C}}$ iff $\eta(\mathscr{C}) < \eta(\hat{\mathscr{C}})$.

## 5. Circuit Manifolds

In this section we provide a detailed proof that a state map (i.e. the map defined by a parameterized quantum circuit according to Definition 1) maps a properly chosen parameter space into a submanifold of $\mathbb{S}^{n-1}$.

### 5.1. State Maps Induce Local Immersions

As introduced in section 2.3.2, let $\mathbb{P} \subseteq \mathbb{R}^k$ be a parameter space and $\mathscr{C}$ be an ansatz depending on the parameters $p_1, \ldots, p_k \in \mathbb{P}$, and let $\Lambda_{\mathscr{C}}$ be the map $\Lambda_{\mathscr{C}} : \mathbb{P} \to \mathbb{S}^{n-1}$ with $p_1, \ldots, p_k \mapsto \mathscr{C}(p_1, \ldots, p_k)|\iota\rangle$ with a chosen fixed initial state $|\iota\rangle$. Thus, $\Lambda_{\mathscr{C}}(\mathbb{P}) \subseteq \mathbb{S}^{n-1}$ is the set of all states reachable by the parameterized circuit $\mathscr{C}$. Lemma 1 shows that $\Lambda_{\mathscr{C}}$ is smooth, i.e. of class $C^s$ for any $s \in \mathbb{N}$.

Assumption 1. From now on, we assume that the parameter space $\mathbb{P} \subseteq \mathbb{R}^k$ is an embedded submanifold of $\mathbb{R}^k$ of dimension $k$. As discussed in section 3.2 and section 3.3, this is often the case in practical situations.

Let $q \in \mathbb{P}$ and let $\text{rank}_q \Lambda_{\mathscr{C}} = r$. Thus, there is an $r \times r$ submatrix of the differential $d(\Lambda_{\mathscr{C}})_q$ whose determinant is not 0; w.l.o.g. with $d(\Lambda_{\mathscr{C}})_q =: (\lambda_{ij})_{1 \leq i,j \leq k}$ this submatrix is $(\lambda_{ij})_{1 \leq i,j \leq r}$. According to Lemma 5, there exists a neighborhood $W \in \mathfrak{U}_q$ open in $\mathbb{R}^k$ such that $\text{rank}_x \Lambda_{\mathscr{C}} \geq r$ for each $x \in W$; w.l.o.g. $W$ is an open ball centered around $q$ contained in $\mathbb{P}$. Next, let $\mathbb{R}_q^r := \{x \in \mathbb{R}^k \mid x_j = q_j, j > r\}$ be the hyperplane of $\mathbb{R}^k$ parallel to $\mathbb{R}^r$ through $q$, and let $\hat{\mathbb{P}} := \mathbb{R}_q^r \cap W$ (see Figure 20); i.e. $\hat{\mathbb{P}}$ is the intersection of $W$ with $\mathbb{R}_q^r$. If $W \subseteq \mathbb{P}$ is chosen small enough it is contained in a chart $(U, \varphi)$ of $\mathbb{P}$ around $q$, i.e. $W \subseteq U$. With $\hat{\mathbb{P}} = \hat{\mathbb{P}} \cap W$ open in $\hat{\mathbb{P}}$ in the subspace topology, $\{(\hat{\mathbb{P}}, \varphi|_{\hat{\mathbb{P}}})\}$ is an atlas of $\hat{\mathbb{P}}$. This shows:

**Note 14:** $\hat{\mathbb{P}} \subseteq \mathbb{P}$ is an embedded submanifold of $\mathbb{P}$ with $\dim \hat{\mathbb{P}} = r$. ∎

It is $\Lambda_{\mathscr{C}}(x_1, \ldots, x_r, \ldots, x_j, \ldots, x_k) = \Lambda_{\mathscr{C}}(x_1, \ldots, x_r, \ldots, \bar{x}_j, \ldots, x_k)$ for $j > r$ and for corresponding points in $\hat{\mathbb{P}}$ (i.e. $x_j = q_j = \bar{x}_j$), thus, $\Lambda_{\mathscr{C}}$ is constant in each of the $x_j$ directions. Consequently, $\partial \Lambda_{\mathscr{C}} / \partial x_j = 0$ for $j > r$. Together with $\mathrm{rank}_x \Lambda_{\mathscr{C}} \geq r$ this implies $\mathrm{rank}_x \Lambda_{\mathscr{C}} = r = \dim \hat{\mathbb{P}}$ for $x \in \hat{\mathbb{P}}$. This proves:

**Lemma 16**: $\hat{\Lambda}_{\mathscr{C}} := \Lambda_{\mathscr{C}}|_{\hat{\mathbb{P}}} : \hat{\mathbb{P}} \to \mathbb{S}^{n-1}$ is an immersion. ∎

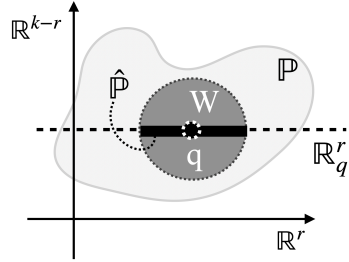

**Fig. 20**. Restricting the Parameter Space

## 5.2. Locally Embedded Circuit Manifolds

According to the Rank Theorem (Theorem 3), there exist charts around $q$ and $f(q)$ such that $\hat{\Lambda}_{\mathscr{C}}$ has the form $\hat{\Lambda}_{\mathscr{C}}(x_1, \ldots, x_k) = (x_1, \ldots, x_r, 0, \ldots, 0)$. I.e. $\hat{\Lambda}_{\mathscr{C}}$ is injective in a neighborhood of $q \in \hat{\mathbb{P}}$. Thus, for $W$ chosen properly (i.e. by possibly shrinking it), $\hat{\Lambda}_{\mathscr{C}}$ is injective, and with $\hat{\Lambda}_{\mathscr{C}}$ being also an immersion, we get (Definition 14):

**Lemma 17**: $\hat{\Lambda}_{\mathscr{C}} = \Lambda_{\mathscr{C}}|_{\hat{\mathbb{P}}}$ is an injective immersion, and $\hat{\Lambda}_{\mathscr{C}}(\hat{\mathbb{P}}) \subseteq \mathbb{S}^{n-1}$ is an immersed submanifold. ∎

According to Note 14, $\hat{\mathbb{P}} \subseteq \mathbb{P}$ is an embedded submanifold of $\mathbb{P}$ with $\dim \hat{\mathbb{P}} = r$. By Assumption 1, $\mathbb{P} \subseteq \mathbb{R}^k$ is an embedded submanifold of $\mathbb{R}^k$. Thus (Lemma 8), both inclusions $\hat{\mathbb{P}} \hookrightarrow \mathbb{P}$ as well as $\mathbb{P} \hookrightarrow \mathbb{R}^k$ are embeddings, which implies that the composed inclusion $\iota : \hat{\mathbb{P}} \hookrightarrow \mathbb{R}^k$ is an embedding (Note 8(e)). According to Lemma 9, the image of the inclusion $\iota(\hat{\mathbb{P}}) = \hat{\mathbb{P}} \subseteq \mathbb{R}^k$ is an embedded submanifold. This proves:

**Note 15:** $\hat{\mathbb{P}} \subseteq \mathbb{R}^k$ is an embedded submanifold with $\dim \hat{\mathbb{P}} = r$. ∎

With $\hat{\mathbb{P}} \subseteq \mathbb{R}^k$ being an embedded submanifold, chose a chart $(U, \varphi)$ of $\hat{\mathbb{P}}$ around $q$, i.e. it is $\varphi(U) \subseteq_{\mathrm{open}} \mathbb{R}^r$. Next, chose a compact ball $B_\varepsilon\left(\varphi(q)\right) \subset_{\mathrm{compact}} \varphi(U)$ with center $\varphi(q)$. Now, $B_\varepsilon\left(\varphi(q)\right)$ is a compact manifold (with boundary) of dimension $r$, and $\varphi^{-1}$ is a diffeomorphism, thus, $M := \varphi^{-1}\left(B_\varepsilon\left(\varphi\left(q\right)\right)\right) \subset \hat{\mathbb{P}}$ is a compact submanifold of dimension $r$. According to Lemma 17, $\hat{\Lambda}_\mathscr{C}$ is an injective immersion, which implies that especially $\hat{\Lambda}_\mathscr{C}|_M : M \to \mathbb{S}^{n-1}$ is an injective immersion. Then, according to Lemma 7(a), $\hat{\Lambda}_\mathscr{C}|_M$ is an embedding, and Lemma 9 finally proves:

**Lemma 18**: $\hat{\Lambda}_\mathscr{C}(M) \subseteq \mathbb{S}^{n-1}$ is an embedded submanifold of dimension $\dim \hat{\Lambda}_\mathscr{C}(M) = \mathrm{rank}_q \Lambda_\mathscr{C}$. ∎

If choosing in the construction before instead of the compact ball $B_\varepsilon\left(\varphi\left(q\right)\right)$ the open ball $\mathring{B}_\varepsilon\left(\varphi\left(q\right)\right)$ it is $U := \varphi^{-1}\left(\mathring{B}_\varepsilon\left(\varphi\left(q\right)\right)\right)$ open in $M$, i.e. $U \in \mathfrak{U}_q$ is an neighborhood of $q$ open in $\hat{\mathbb{P}}$, i.e. $q \in U \subseteq_{\mathrm{open}} M$. $\hat{\Lambda}_\mathscr{C}|_M$ is an embedding, especially a homeomorphism onto the image, i.e. $\hat{\Lambda}_\mathscr{C}(U) \subseteq_{\mathrm{open}} \hat{\Lambda}_\mathscr{C}(M)$. The lemma before showed that $\hat{\Lambda}_\mathscr{C}(M) \subseteq \mathbb{S}^{n-1}$ is an embedded submanifold.

Moreover, an open set $O \subseteq_{\mathrm{open}} N$ of a manifold $N$ is again a manifold with $\dim O = \dim N$ (Note 5(b)). The inclusion $\iota : O \hookrightarrow N$ is the identity in proper charts $(U, \varphi)$ of N and corresponding charts $(O, \varphi|_O)$ of $O$ (w.l.o.g $O \subseteq U$), and thus, it is an immersion. With $O \subseteq_{\mathrm{open}} N$ the inclusion is also a homeomorphism onto $O = \iota(O)$. In summary, $\iota : O \hookrightarrow N$ is an embedding.

With $\hat{\Lambda}_\mathscr{C}(U) \subseteq_{\mathrm{open}} \hat{\Lambda}_\mathscr{C}(M)$, the latter proves that $\hat{\Lambda}_\mathscr{C}(U)$ is an embedded submanifold of the embedded submanifold $\hat{\Lambda}_\mathscr{C}(M) \subseteq \mathbb{S}^{n-1}$ (Lemma 9). As a consequence, $\hat{\Lambda}_\mathscr{C}(U)$ is an embedded submanifold of $\mathbb{S}^{n-1}$(Note 8(e)). Furthermore, $\dim \hat{\Lambda}_\mathscr{C}(U) = \dim \hat{\Lambda}_\mathscr{C}(M) = \mathrm{rank}_q \hat{\Lambda}_\mathscr{C}$. Since the overall construction is valid for each $p \in \hat{\mathbb{P}}$ the following has been proven:

**Note 16:** For each $p \in \hat{\mathbb{P}}$ there exists an open neighborhood of $p$, i.e. $p \in U \subseteq_{\mathrm{open}} \hat{\mathbb{P}}$, such that $\hat{\Lambda}_\mathscr{C}|_U$ is an embedding, and $\hat{\Lambda}_\mathscr{C}(U) \subseteq \mathbb{S}^{n-1}$ is an embedded submanifold with $\dim \hat{\Lambda}_\mathscr{C}(U) = \mathrm{rank}_q \hat{\Lambda}_\mathscr{C}$. ∎

Both, $U$ and $\hat{\Lambda}_\mathscr{C}(U)$ are differentiable manifolds without boundary of the same dimension. Moreover, by construction $\hat{\Lambda}_\mathscr{C}|_U : U \to \hat{\Lambda}_\mathscr{C}(U)$ is differentiable with bijective differential $d\left(\hat{\Lambda}_\mathscr{C}|_U\right)_p$. The Inverse Function Theorem (Theorem 2) then implies:

**Note 17:** For each $p \in \hat{\mathbb{P}}$ there exists an open neighborhood $U_p$ of $p$, i.e. $p \in U_p \subseteq_{\text{open}} \hat{\mathbb{P}}$, such that $\hat{\Lambda}_{\mathscr{C}}|_{U_p} : U_p \to \hat{\Lambda}_{\mathscr{C}}(U_p)$ is a diffeomorphism. ∎

### 5.3. An Attempt to Extend Locally Embedded Circuit Manifolds

Many local topological properties, i.e. properties valid in a neighborhood of a point, can be extended to connected components. For example, a function that is locally constant, is constant on each connected component. Central for corresponding proofs is Lemma 2. In the case of locally constant functions, its use is as follows:

Let $X$ be a connected topological space and let $f : X \to \mathbb{R}$ be locally constant, i.e. each point $x \in X$ has an open neighborhood $x \in U_x \subseteq_{\text{open}} X$ and a number $c_x \in \mathbb{R}$ such that $f|_{U_x} = c_x$. Furthermore, if two such neighborhoods $U_x, U_y$ intersect, the corresponding numbers $c_x, c_y$ are equal: for $z \in U_x \cap U_y$ it is $f(z) = c_x$ and $f(z) = c_y$, i.e. $c_x = c_y$ and, thus, $f|_{U_x} = f|_{U_y}$. Let $\mathscr{U} = \{U_x \mid x \in X\}$ be an open cover consisting of such neighborhoods, and choose two arbitrary points $a, b \in X$. According to Lemma 2, there is an open chain $\{U_1, \ldots, U_n\} \subseteq \mathscr{U}$ connecting $a$ and $b$. Because $U_i \cap U_{i+1} \neq \emptyset$ it is $f|_{U_i} = f|_{U_{i+1}}$, which implies that $f|_{U_i} = f|_{U_j}$ for any $1 \leq i, j \leq n$. Thus, $f|_{U_1} = f|_{U_n}$, i.e. $f(a) = f(b)$. Because $a, b$ are arbitrary points from $X$, $f$ is constant.

In our context, the question at hand is whether the rank of a differentiable map (which can locally only increase according to Lemma 5) can globally only increase on a connected component. In order to mimic the proof before, we make the following assumption on the parameter space.

<u>Assumption 2</u>: For each $q \in \mathbb{P}$ and each $1 \leq r \leq k$ let $\overline{\mathbb{P}}_{q,r} := \mathbb{R}_q^r \cap \mathbb{P}$ be connected. I.e. especially, $q \in \overline{\mathbb{P}}_{q,r}$.

More precisely, it suffice that $\overline{\mathbb{P}}_{q,r}$ has a finite number of connected components, and for the construction that follows the connected component containing $q$ is chosen. Thus, w.l.o.g. we can assume that $\overline{\mathbb{P}}_{q,r}$ is connected. This is directly the case in many practical situations, e.g. if $\mathbb{P}$ is an appropriate (see the discussion at the end of section 4.2) semi-open hypercube (see Figure 21). Similarly, a $k$-dimensional torus is an example of a parameter space $\mathbb{P}$ where $\overline{\mathbb{P}}_{q,r}$ consist of more than one but a finite number of connected components.

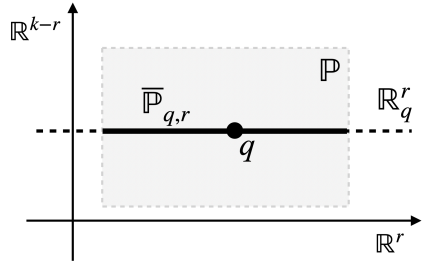

**Fig. 21.** Slicing the Parameter Space

Let $q \in \overline{\mathbb{P}}_{q,r_q}$ with $r_q := \text{rank}_q \Lambda_{\mathscr{C}}$. According to the construction of section 5.1, there exists an open neighborhood of $q$ with $q \in U_q \subseteq_{\text{open}} \overline{\mathbb{P}}_{q,r_q}$ such that $\Lambda_{\mathscr{C}}|_{U_q}$ is an immersion (Lemma 16), i.e. $\Lambda_{\mathscr{C}}|_{U_q}$ has constant rank $r_q$ in $U_q$.

For $x, y \in \overline{\mathbb{P}}_{q,r_q}$ choose $U_x, U_y \subseteq \mathbb{P}$ as before such that $\Lambda_{\mathscr{C}}|_{U_x}$ has constant rank $r_x$ in $U_x$ and that that $\Lambda_{\mathscr{C}}|_{U_y}$ has constant rank $r_y$ in $U_y$. Note that $x \in U_x \subseteq_{\text{open}} \overline{\mathbb{P}}_{x,r_x}$ but not necessarily $U_x \subseteq_{\text{open}} \overline{\mathbb{P}}_{q,r_q}$ or $U_x \subseteq_{\text{open}} \mathbb{P}$; similarly for $U_y$. Figure 22 depicts this situation where $\overline{\mathbb{P}}_{q,r_q}$ is two dimensional, $U_x$ one dimensional, and $U_y$ three dimensional. Thus, it can not be guaranteed that an open cover of $\overline{\mathbb{P}}_{q,r_q}$ exists for which the rank of $\Lambda_{\mathscr{C}}|_U$ is constant for a member $U$ of the open cover. Consequently, even if a $\overline{\mathbb{P}}_{q,r}$ is connected it can not be guaranteed via an argument involving Lemma 2 that the rank is constant on all of $\overline{\mathbb{P}}_{q,r}$.

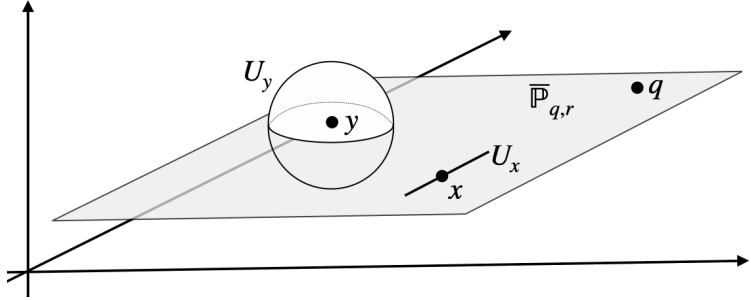

**Fig. 22**. Neighborhoods of Constant Rank of $\Lambda_{\mathscr{C}}$

There is even a counter example:

Let $f : \mathbb{R}^2 \to \mathbb{R}^2, (x, y) \mapsto (x^2, y^2)$. Then $df_{(x,y)} = \begin{pmatrix} 2x & 0 \\ 0 & 2y \end{pmatrix}$. The rank of $df_{(x,y)}$ is as follows:

- Let $(x, y) = (0,0)$. Then $df_{(0,0)} = \begin{pmatrix} 0 & 0 \\ 0 & 0 \end{pmatrix} \Rightarrow \text{rank}_{(0,0)} f = 0$

- Let $y = 0 \wedge x \neq 0$. Then $df_{(x,0)} = \begin{pmatrix} 2x & 0 \\ 0 & 0 \end{pmatrix} \Rightarrow \text{rank}_{(x,0)} f = 1$

- Let $y \neq 0 \wedge x = 0$. Then $df_{(0,y)} = \begin{pmatrix} 0 & 0 \\ 0 & 2y \end{pmatrix} \Rightarrow \text{rank}_{(0,y)} f = 1$

- Let $y \neq 0 \wedge x \neq 0$. Then $df_{(x,y)} = \begin{pmatrix} 2x & 0 \\ 0 & 2y \end{pmatrix} \Rightarrow \text{rank}_{(x,y)} f = 1$

The following figure shows the landscape of ranks of this map $f$.

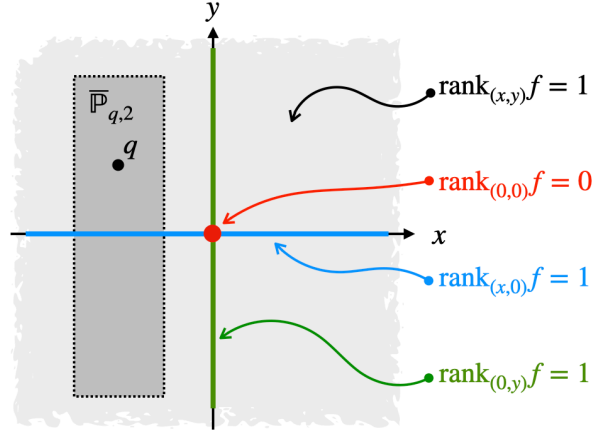

**Fig. 23**. Landscape of Ranks of $f(x, y) = (x^2, y^2)$

$f$ has rank 0 at the origin (red dot), and rank 1 at the x-axis and the y-axis (the green and blue lines without the origin). Each point of the rest of the plain (grey shaded area) has rank 2. For the point $q$ shown, it is $\mathrm{rank}_q f = 2 = r_q$. With a properly chosen parameter space $\mathbb{P}$ (e.g. the dark grey shaded rectangle), the set $\overline{\mathbb{P}}_{q,2}$ is as depicted. $\overline{\mathbb{P}}_{q,2}$ is a manifold without boundary, $\overline{\mathbb{P}}_{q,2}$ is connected, and $\dim \overline{\mathbb{P}}_{q,2} = 2$; i.e. $\overline{\mathbb{P}}_{q,2}$ satisfies Assumptions 1 and 2. However, the rank of $f$ is not constant on $\overline{\mathbb{P}}_{q,2}$.

### 5.4. Determining "Large" Circuit Manifolds

The counter example before (and Figure 23) indicates that within a set $\overline{\mathbb{P}}_{q,r_q}$ "large" areas $A$ exist in which the rank of $\mathrm{rank} \Lambda_{\mathscr{C}}|_A = r_q$ is constant, i.e. $\Lambda_{\mathscr{C}}|_A$ is an immersion. For example, $\mathscr{U} = \{U_x \subseteq_{\mathrm{open}} \overline{\mathbb{P}}_{q,r_q} \mid \mathrm{rank} \Lambda_{\mathscr{C}}|_{U_x} = r_q\} \neq \emptyset$ and $A := \bigcup_{U \in \mathscr{U}} U$.

Locally, $\Lambda_{\mathscr{C}}|_A$ is also injective (see the proof of Lemma 17). Thus, $\Lambda_{\mathscr{C}}(A)$ is locally an immersed submanifold of $\mathbb{S}^{n-1}$. The domain of injectivity can be extended in concrete situations by analyzing $\Lambda_{\mathscr{C}}(A)$. Here is an example of such an analysis:

<u>Example</u>: Let $\mathbb{P} = \mathbb{R}$ and let $\mathscr{C}(x) = e^{ixH}$ for a hermitian matrix $H$. A set $A \subseteq \mathbb{R}$ has to be determined such that $\mathscr{C}(x)$ is injective. Since every hermitian matrix is normal by definition ([LM15] Definition 18.1) and each normal matrix is diagonalizable ([LM15] Theorem 18.2), it is $H = T \mathrm{diag}\left(e^{i\lambda_1}, \cdots, e^{i\lambda_n}\right) T^*$ for a unitary matrix $U$ and the real eigenvalues $\lambda_1, \ldots, \lambda_n$. Consequently, it is $e^{ixH} = T \mathrm{diag}\left(e^{ix\lambda_1}, \cdots, e^{ix\lambda_n}\right) T^*$. Finding the set of parameters for which $\mathscr{C}(x) = e^{ixH}$ is injective means to determine when $e^{ixH} = e^{iyH}$ implies $x = y$. Thus, we need to find $X \subseteq \mathbb{R}$ such that $e^{ixH} = e^{iyH} \Leftrightarrow x = y$ for $x, y \in X$.

diag$\left(e^{ix\lambda_1}, \cdots, e^{ix\lambda_n}\right)$, and thus $e^{ixH}$, may be periodic - which implies non-injectivity: If $\lambda_1, \ldots, \lambda_n \in \mathbb{Q}$, then there exists a $T > 0$ such that $e^{iHx} = e^{iH(x+T)}$ for each $x$.

Assume $x \neq y$ with $e^{ixH} = e^{iyH}$. Then, $e^{i(x-y)H} = I$. Thus, the spectrum of $(x-y)H$ must be in $2\pi\mathbb{Z}$, i.e. $\lambda_j(x-y) \in 2\pi\mathbb{Z}$ for each eigenvalue $\lambda_j$ of $H$. If the spectrum of $H$ fulfills

$$\forall\, T > 0\ :\ \sigma(H) = \{\lambda_1, \ldots, \lambda_n\} \nsubseteq \frac{2\pi}{T}\mathbb{Z} \tag{15}$$

then $e^{i(x-y)H} \neq I$, i.e. $e^{ixH} \neq e^{iyH}$: $\mathscr{C}(x)$ is injective. Thus, if $H$ fulfills condition (15), $\mathscr{C}$ and thus $\Lambda_\mathscr{C}$ is injective on all of $\mathbb{R} = \mathbb{P}$. ∎

In summary, for $A := \bigcup_{U \in \mathscr{U}} U \subseteq \overline{\mathbb{P}}_{q,r_q}$ one has to determine in each case of $\mathscr{C}$ the subset $G \subseteq A$ (the "good set") of $A$ in which $\Lambda_\mathscr{C}$ is injective. Then, $\Lambda_\mathscr{C}(G)$ is locally an immersed submanifold of $\mathbb{S}^{n-1}$. By choosing a "large" compact subset $K \subseteq_{\text{compact}} G$ the embedding submanifold $\Lambda_\mathscr{C}(K) \subseteq \mathbb{S}^{n-1}$ results (Lemma 7, Lemma 9).

## 6. Conclusion

The literature about expressivity of parameterized quantum circuits requires a lot of background in differential topology that makes it hard to comprehend. This contribution provides the corresponding background in a single place with detailed references to textbooks and seminal papers which may be consulted for a much deeper dive into the domain. Similarly, statements about properties of parameterized quantum circuits are often not proved in the literature or a proof is only indicated. This contribution provides proves of statement about dimensional expressivity (to be precise: proves for local versions of such properties). Also, counter-examples are given, limits are pointed out (e.g. by highlighting the importance of singularities).

Section 5 clearly reveals the need for future work: Conditions under which local embeddings can be extended to (global) embedding in the case of parameterized quantum circuits are needed, especially if the circuits have the form of equation (5) which is practically relevant. Furthermore, the applicability in practical situations of the proceeding sketched in Section 5.4 must be evaluated in practice.

**Author Contributions**: Writing – original draft, F.L.; Writing – review & editing, J.B. All authors have read and agreed to the published version of the manuscript.

## Appendix A

At the end of section 4.5 it is shown that $df_p(v) = Df(p)(v) = D_v f(p)$, i.e. the differential applied to $v$ is the directional derivative of $f$ in the direction of $v$. By definition, the directional derivative is $D_v f(p) := \lim_{t \to 0} \frac{f(p+tv) - f(p)}{t}$, i.e. $df_p(v) = D_v f(p) = \lim_{t \to 0} \frac{f(p+tv) - f(p)}{t}$.

Thus, for $f(X) := X^*X$ we get:

$$\begin{aligned} df_A(V) &= \lim_{t \to 0} \frac{f(A+tV) - f(A)}{t} \\ &= \lim_{t \to 0} \frac{(A+tV)^*(A+tV) - A^*A}{t} \\ &= \lim_{t \to 0} \frac{(A^* + tV^*)(A+tV) - A^*A}{t} \\ &= \lim_{t \to 0} \frac{A^*A + tA^*V + tV^*A + t^2 V^*V - A^*A}{t} \\ &= \lim_{t \to 0} \frac{t(A^*V + V^*A + tV^*V)}{t} \\ &= \lim_{t \to 0} (A^*V + V^*A + tV^*V) \\ &= A^*V + V^*A \end{aligned}$$

This proves the claim $df_A(V) = A^*V + V^*A$.

## Appendix B

We sketch a proof that $f : ]-\pi, +\pi[ \to \mathbb{R}^2$, $x \mapsto (\sin 2x, \sin x)$ is injective. Assume $s, t \in ]-\pi, +\pi[$ and $f(s) = f(t)$. Thus:

$$\begin{aligned} &\sin 2s = \sin 2t \;\wedge\; \sin s = \sin t \\ &\stackrel{(A)}{\Leftrightarrow} 2 \sin s \cos s = 2 \sin t \cos t \;\wedge\; \sin s = \sin t \\ &\stackrel{(B)}{\Leftrightarrow} \sin s \cos s = \sin t \cos t \;\wedge\; \sin s = \sin t \\ &\stackrel{(C)}{\Leftrightarrow} \cos s = \cos t \;\wedge\; \sin s = \sin t \end{aligned}$$

(A) is because of $\sin 2x = 2 \sin x \cos x$, (B) is the division by "2", and (C) is because of $\sin s = \sin t$ and division by $\sin s$, w.l.o.g. $\sin s \neq 0$ (otherwise $s = t = 0$ because of $s, t \in ]-\pi, +\pi[$, which implies injectivity).

Next, injectivity of both, $\sin : ]-\pi/2, +\pi/2[ \to \mathbb{R}$ and $\cos : ]0, +\pi[ \to \mathbb{R}$ is considered, thus: $\cos s = \cos t$ for $t \in ]0, \pi[$ implies that $s \in ]-\pi, 0]$; and vice versa $t \in ]-\pi, 0[$ implies that $s \in [0, \pi[$. Otherwise $\sin s = \sin t$ for $t \in ]-\pi/2, +\pi/2[$ implies $s \geq \pi/2$ or $s \leq -\pi/2$.

Now, a case distinction is performed. Case 1: $t \in ]0, \pi/2[$ implies both, $s \in ]-\pi, 0]$ as well as $s \leq -\pi/2$. But for $t \in ]0, \pi/2[$ and $s \in ]-\pi, -\pi/2[$ it is $\sin t \neq \sin s$. The other cases follow analogously.